%% file: main.tex
\documentclass[english,aps, reprint, prb, amsmath,amssymb,floatfix,superscriptaddress,eqsecnum,longbibliography]{revtex4-2}
\usepackage[T1]{fontenc}
\usepackage{tabularx}
\usepackage{babel}
\usepackage{amsmath}
\usepackage{graphicx}
\usepackage{units}
\usepackage[unicode=true,pdfusetitle,
 bookmarks=true,bookmarksnumbered=false,bookmarksopen=false,
 breaklinks=false,pdfborder={0 0 1},backref=false,colorlinks=false]
 {hyperref}

\makeatletter
\usepackage{braket}

\makeatother

\begin{document}
\title{Adiabatic spin dynamics and effective exchange interactions from constrained
tight-binding electronic structure theory: Beyond the Heisenberg regime}
\author{Simon Streib}
\affiliation{Department of Physics and Astronomy, Uppsala University, Box 516,
SE-75120 Uppsala, Sweden}
\author{Ramon Cardias}
\affiliation{Department of Applied Physics, School of Engineering Sciences, KTH
Royal Institute of Technology, Electrum 229, SE-16440 Kista, Sweden}
\author{Manuel Pereiro }
\affiliation{Department of Physics and Astronomy, Uppsala University, Box 516,
SE-75120 Uppsala, Sweden}
\author{Anders Bergman}
\affiliation{Department of Physics and Astronomy, Uppsala University, Box 516,
SE-75120 Uppsala, Sweden}
\author{Erik Sj\"oqvist }
\affiliation{Department of Physics and Astronomy, Uppsala University, Box 516,
SE-75120 Uppsala, Sweden}
\author{Cyrille Barreteau}
\affiliation{SPEC, CEA, CNRS, Universit\'e Paris-Saclay, CEA Saclay, 91191 Cedex Gif-sur-Yvette, France}
\author{Anna Delin }
\affiliation{Department of Applied Physics, School of Engineering Sciences, KTH
Royal Institute of Technology, Electrum 229, SE-16440 Kista, Sweden}
\author{Olle Eriksson }
\affiliation{Department of Physics and Astronomy, Uppsala University, Box 516,
SE-75120 Uppsala, Sweden}
\affiliation{School of Science and Technology, \"Orebro University. SE-70182 \"Orebro,
Sweden}
\author{Danny Thonig}
\affiliation{School of Science and Technology, \"Orebro University. SE-70182 \"Orebro,
Sweden}
\affiliation{Department of Physics and Astronomy, Uppsala University, Box 516,
SE-75120 Uppsala, Sweden}
\date{June 7, 2022}

\begin{abstract}
We consider an implementation of the adiabatic spin dynamics approach
in a tight-binding description of the electronic structure. The adiabatic
approximation for spin-degrees of freedom assumes that the faster
electronic degrees of freedom are always in a quasi-equilibrium state,
which significantly reduces the numerical complexity in comparison to 
the full electron dynamics. Non-collinear
magnetic configurations are stabilized by a constraining field, which
allows to directly obtain the effective magnetic field from the
negative of the constraining field. While the dynamics
are shown to conserve energy, we demonstrate that adiabatic
spin dynamics does not conserve the total spin angular momentum when
the lengths of the magnetic moments are allowed to change, which is 
confirmed by numerical simulations. Furthermore, we develop a method to 
extract an effective two-spin exchange interaction from the energy curvature tensor of 
non-collinear states, which we calculate at each time step of the numerical simulations. We demonstrate the effect of non-collinearity on 
this effective exchange and limitations due to multi-spin interactions in strongly non-collinear
configurations beyond the regime where the Heisenberg model is valid. The relevance of the results are discussed with respect to experimental pump-probe experiments that follow the ultra-fast dynamics of magnetism.
\end{abstract}
\maketitle

\section{Introduction}

Adiabatic spin dynamics \cite{Antropov1995,Antropov1996, Halilov1998}
is based on the assumption that the spin-degrees of freedom, corresponding
to the direction of magnetic moment vectors, are much slower than
the electronic degrees of freedom, which are related to changes of
the lengths of magnetic moments. The effective field that drives the
spin dynamics is within this approach obtained from the gradient of
the electronic energy with respect to the moment directions, where
the electronic system is considered to be in its ground state for
a given moment configuration. Since an arbitrary moment configuration
does not correspond to the absolute ground state, non-equilibrium configurations
need to be stabilized. This can be done exactly by introducing constraining
fields \cite{Stocks1998,Ujfalussy1999,Ma2015} or approximately by fixing
the quantization axis for each moment, which does not align the moments
exactly \cite{Singer2005}. The main advantages of
adiabatic spin dynamics compared to the standard atomistic spin dynamics
approaches \cite{Bertotti2009,Eriksson2017} are that it does not
require a parametrization of a classical spin model, such as the Heisenberg
model, and it does not assume constant magnetic moment lengths and
exchange parameters. Dynamical changes of these parameters
could be important for strongly non-collinear states of itinerant magnets, which can occur, e.g., after
ultrafast demagnetization by a laser pulse \citep{Beaurepaire1996} and when approaching the Curie temperature of the magnet \cite{Boettcher2012,Mankovsky2013,Mankovsky2020b} 
The main disadvantage is the increased
numerical complexity since at each time step of the dynamics the corresponding
electronic ground state needs to be calculated, which is much more
demanding than dealing with a classical spin model with constant parameters.
Therefore, despite
its advantages and conceptual elegance, adiabatic spin dynamics has
not been widely adopted so far.  One example is an application of adiabatic
spin dynamics without constraining fields for a chain of ten Co atoms on an Au(001)
surface \cite{Rozsa2014}. Within their implementation \cite{Rozsa2014},
the electronic structure is only calculated once with density functional
theory (DFT) for the ground state and non-equilibrium configurations are
implemented without self-consistency by only rotating the exchange
field, which reduces the numerical complexity considerably but can
only expected to be reliable close to the ground state configuration
where a standard atomistic spin dynamics description would also be sufficient.

In recent works, the precise relation between the effective field,
energy gradient, and constraining field have been established in the
context of tight-binding and DFT \cite{Streib2020},
and adiabatic spin dynamics simulations have been performed within
a completely self-consistent tight-binding description with constraining
fields for Fe, Co and Ni dimers \cite{Cardias2021a}, where agreement
with a simple Heisenberg model was found for small angles between
the two magnetic moments of the dimer.  In this work, we consider spin dynamics
for strongly non-collinear configurations beyond the Heisenberg regime and we address
two fundamental aspects of adiabatic spin dynamics: the conservation
of energy and angular momentum. We show both analytically and with
spin dynamics simulations that, while energy is conserved as expected,
angular momentum is not conserved due to dynamical changes in the lengths of
magnetic moments. 

When considering non-collinear states, the question
arises how the exchange interaction $J_{ij}$ between two spins is
affected in comparison to the ground state. Recently, 
the concept of a local spin Hamiltonian was introduced \cite{Streib2021}, which
is based on the energy curvature tensor $\mathcal{J}_{ij}^{\alpha\beta}$ and accurately
describes small fluctuations around a reference moment configuration.
While $\mathcal{J}_{ij}^{\alpha\beta}$ gives full access
to the energy curvature with respect to rotations of the spins at
sites $i$ and $j$ in directions $\alpha,\beta=x,y,z$, it is a quantity
that is not straight-forward to interpret in non-collinear states,
even for a simple Heisenberg model. This complication arises from the fact
that infinitesimal rotations of magnetic moments correspond only to the perpendicular
component to the gradient in Cartesian coordinates with respect to the moment directions.
It should be noted however, that despite these difficulties, there are suggestions on how to evaluate exchange interactions from non-collinear states \citep{Antropov1997,Antropov1999,Szilva2013,Secchi2015,Szilva2017,Cardias2020a,Cardias2020b}.
We demonstrate here how the curvature tensor is related to a general
two-spin isotropic exchange interaction and show how an effective exchange interaction
$J_{ij}$ can be obtained. Although this
effective exchange does not contain the full curvature information,
it is a useful quantity to characterize the effect of non-collinearity
on the exchange interaction, indicating non-Heisenberg behavior. We
demonstrate this by spin dynamics simulations of Fe and Co spin chains
with a phenomenological Gilbert damping, which allows us to track
the effective exchange during the relaxation process back to the 
ground state. While we find that the average nearest-neighbor exchange
interaction is increased by about $10\;\%$ in our initial random
configuration, individual exchange interactions fluctuate very strongly,
which could have an impact on accurately determining the critical
temperature of a magnetic material based on spin dynamics simulations.
This should be contrasted with disordered local moment (DLM) calculations
of exchange parameters \cite{Boettcher2012,Mankovsky2013,Mankovsky2020b},
which provide only an average change of the exchange interaction due
to spin disorder and do not capture the strong fluctuations that we
observe. Furthermore, we argue that multi-spin interactions \cite{Tanaka1977,MacDonald1988,Singer2011,Hoffmann2020,Brinker2020,Mankovsky2020}
limit the reliability of a two-spin exchange model in strongly non-collinear
configurations, as indicated by our numerical calculations.

The paper is structured as follows: in Sec.~\ref{sec:Adiabatic-spin-dynamics},
we discuss adiabatic spin dynamics and introduce the equation of motion
based on the constraining field. We derive the conservation of energy
in Sec.~\ref{sec:Conservation-laws} and show that angular momentum
is not conserved when magnetic moment lengths are not constant. We
introduce in Sec.~\ref{sec:Effective-Heisenberg-exchange} an effective
exchange interaction and show how it can be calculated from the curvature
tensor. In Sec.~\ref{sec:Tight-binding-model}, we discuss the tight-binding
electronic structure description with a special focus on the magnetic
Stoner contribution and the calculation of the energy curvature tensor. We
apply this formalism in Sec.~\ref{sec:Dynamics} to Fe, Co and Ni
dimers and Fe and Co spin chains, which allows to study their
dynamics and test theoretical predictions. We summarize the results
and discuss their consequences for the adiabatic spin dynamics framework
in Sec.~\ref{sec:Summary-and-discussion}. In Appendix~\ref{sec:Derivative-matrix},
we provide the derivative matrix for rotations of magnetic moments
in Cartesian coordinates. The contribution of a Dzyaloshinskii--Moriya
interaction to the energy curvature tensor is given in Appendix~\ref{sec:DMI}.

\section{Adiabatic spin dynamics\label{sec:Adiabatic-spin-dynamics}}

Adiabatic spin dynamics is based on the assumption that electronic
degrees of freedom are much faster than the dynamics of the magnetic
moment directions $\lbrace\mathbf{e}_{i}\rbrace$ \cite{Antropov1995,Antropov1996,Halilov1998}, where $i$ denotes the lattice site. This assumption is rigorously justified for spin-wave excitations with an energy much smaller than the Stoner spin splitting \cite{Antropov2003,Katsnelson2004}. Deviations between adiabatic spin-wave spectra and non-adiabatic spectra based on the transverse dynamic magnetic susceptibility obtained from time-dependent DFT have been found for high-energy spin waves \cite{Buczek2011,Durhuus2022}. In the spin dynamics simulations that we consider here, we are dealing with time scales above $1\;\mathrm{fs}$, whereas the relevant electronic relaxation time is below $1\;\mathrm{fs}$, which can be estimated from the electron band width \cite{LandauLifshitz3,Toews2015}, supporting the application of the adiabatic approximation. However, for a complete theoretical description of the ultrafast demagnetization by a laser pulse \cite{Beaurepaire1996}, it is important to take the electron dynamics into account and go beyond the adiabatic approximation to describe the initial laser-induced excitation of electrons \cite{Toews2015}.

Within the adiabatic approximation, the energy depends only on the moment directions,
\begin{equation}
E =E(\{\mathbf{e}_{i}\}),\label{eq:adiabatic energy}
\end{equation}
and the electronic degrees of freedom can be considered to be in a
quasi-equilibrium state with fixed moment directions and relaxed magnetic moment lengths
such that the energy is minimal with respect to the moment lengths. For the calculation
of this electronic state, it is necessary to constrain the moment
directions to point along the required directions $\{\mathbf{e}_{i}\}$,
as otherwise the system would relax back to the absolute ground state
\cite{Stocks1998,Ujfalussy1999}.

We implement the constraint on the moment directions by adding a constraining
field $\mathbf{B}_{i}^{\text{con}}$ to the electronic tight-binding
Hamiltonian, $\hat{\mathcal{H}}_{\mathrm{tb}}$, 
\begin{equation}
\hat{\mathcal{H}}=\hat{\mathcal{H}}_{\mathrm{tb}}+\hat{\mathcal{H}}_{\text{con}},
\end{equation}
with 
\begin{equation}
\hat{\mathcal{H}}_{\text{con}}=-\sum_{i}\hat{\mathbf{M}}_{i}\cdot\mathbf{B}_{i}^{\text{con}},
\end{equation}
where $\hat{\mathbf{M}}_{i}$ is the total magnetic moment operator
at lattice site $i$. The constraining field is designed to be perpendicular to the
moment directions, $\mathbf{B}_{i}^{\text{con}}\cdot\mathbf{e}_{i}=0$,
and constrains therefore only the moment directions and not their lengths $M_{i}$.
We employ the following iterative algorithm for calculating the constraining
field \cite{Stocks1998,Ujfalussy1999}, 
\begin{align}
\mathbf{B}_{i}^{\text{con}}(k+1) & =\mathbf{B}_{i}^{\text{con}}(k)-\left(\mathbf{B}_{i}^{\text{con}}(k)\cdot\mathbf{e}_{i}\right)\mathbf{e}_{i}\nonumber \\
 & -B_{0}\left[\mathbf{m}_{i}-\left(\mathbf{m}_{i}\cdot\mathbf{e}_{i}\right)\mathbf{e}_{i}\right],\label{eq:Stocks}
\end{align}
where $k$ is the iteration index, $B_{0}$ a free parameter that
can be tuned for optimal convergence, and $\mathbf{m}_{i}=\langle\hat{\mathbf{M}}_{i}\rangle/M_{i}$
is the output moment direction from the electronic structure calculation.

Within this constrained tight-binding approach, the effective field
acting on a magnetic moment is given by \cite{Streib2020}
\begin{equation}
\mathbf{B}_{i}^{\text{eff}}=-\mathbf{B}_{i}^{\text{con}}.
\end{equation}
The equation of motion of the moment directions at zero temperature is \cite{Antropov1995,Antropov1996,Eriksson2017}
\begin{align}
\dot{\mathbf{e}}_{i} & =\frac{\gamma}{1+\alpha^2}\mathbf{e}_{i}\times\mathbf{B}_{i}^{\text{eff}}+\frac{\alpha\gamma}{1+\alpha^2}\mathbf{e}_{i}\times\left(\mathbf{e}_{i}\times\mathbf{B}_{i}^{\text{eff}}\right),
\end{align}
where $\gamma=-g\mu_{B}/\hbar$ and we allow for a phenomenological
Gilbert damping $\alpha$ \cite{Gilbert2004}. For the numerical integration of this equation of motion, we use the implicit midpoint method, see Ref.~\cite{Mentink2010} for a comparison of integration methods. In this manuscript, our focus is on the
effective field and we refer for the ab initio determination of the
damping parameter to the review in Ref.~\cite{Eriksson2017}.
For the effects of non-collinearity on damping, see for example Refs.~\cite{Yuan2014,Mankovsky2018,Brinker2022}.

\section{Conservation laws\label{sec:Conservation-laws}}

A classical spin Hamiltonian of the typical Heisenberg form,
\begin{equation}
\mathcal{H}_{s}=-\frac{1}{2}\sum_{ij}J_{ij}\mathbf{e}_{i}\cdot\mathbf{e}_{j},
\end{equation}
with Heisenberg exchange $J_{ij}=J_{ji}$, conserves both
energy and total spin angular momentum. In the following, we discuss
the conservation of energy and angular momentum in the context of
adiabatic spin dynamics.

\subsection{Energy conservation}

The energy within the adiabatic approximation depends only on the
instantaneous magnetic configuration $\{\mathbf{e}_{i}\}$. After each time step in the spin dynamics, the system is allowed to
relax back to the quasi-equilibrium state corresponding to the configuration
$\{\mathbf{e}_{i}\}$. Since this relaxation implicitly includes
a coupling to a bath, the system is not closed and the
question arises if and under which conditions energy is conserved.

The time derivative of the energy is given by
\begin{equation}
\dot{E}=\sum_{i\alpha}\frac{\partial E}{\partial e_{i\alpha}}\dot{e}_{i\alpha}=\sum_{i}\boldsymbol{\nabla}_{\mathbf{e}_{i}}E\cdot\dot{\mathbf{e}}_{i}.
\end{equation}
Using the equation of motion (without damping),
\begin{align}
\dot{\mathbf{e}}_{i} & =\gamma\mathbf{e}_{i}\times\mathbf{B}_{i}^{\text{eff}},\label{eq:equation of motion}
\end{align}
we obtain
\begin{align}
\dot{E} & =\sum_{i}\boldsymbol{\nabla}_{\mathbf{e}_{i}}E\cdot\left(\gamma\mathbf{e}_{i}\times\mathbf{B}_{i}^{\text{eff}}\right)\nonumber \\
 & =\sum_{i}\gamma\left[\mathbf{B}_{i}^{\text{eff}}\times\boldsymbol{\nabla}_{\mathbf{e}_{i}}E\right]\cdot\mathbf{e}_{i}.
\end{align}
Therefore, energy is conserved when the effective magnetic field is
proportional to the energy gradient, i.e.,
\begin{equation}
\mathbf{B}_{i}^{\text{eff}}\propto-\frac{1}{M_{i}}\boldsymbol{\nabla}_{\mathbf{e}_{i}}E.
\end{equation}
The effective field in a tight-binding model is given by the negative of the constraining
field and can be related to the energy gradient at zero temperature via the constraining field theorem \cite{Streib2020},
\begin{equation}
-\frac{1}{M_{i}}\boldsymbol{\nabla}_{\mathbf{e}_{i}}E=-\mathbf{B}_{i}^{\text{con}}-\frac{1}{M_{i}}\left\langle \boldsymbol{\nabla}_{\mathbf{e}_{i}}\hat{\mathcal{H}}_\text{tb}\right\rangle,
\end{equation}
where the expectation value is taken with respect to the constrained electronic ground state.
Therefore, energy is conserved if the last term above vanishes,
\begin{equation}
\frac{1}{M_{i}}\left\langle \boldsymbol{\nabla}_{\mathbf{e}_{i}}\hat{\mathcal{H}}_\text{tb}\right\rangle =0,
\end{equation}
such that
\begin{equation}
\mathbf{B}_{i}^{\text{eff}}=-\mathbf{B}_{i}^{\text{con}}=-\frac{1}{M_{i}}\boldsymbol{\nabla}_{\mathbf{e}_{i}}E.
\end{equation}

While for a fundamental Hamiltonian the quantity $\langle \boldsymbol{\nabla}_{\mathbf{e}_{i}}\hat{\mathcal{H}}_\text{tb}\rangle$ vanishes and energy is conserved since there is no explicit dependence on the moment directions, such a dependence may arise in a mean-field description, which we discuss in Sec.~\ref{sec:Tight-binding-model}.

\subsection{Angular momentum conservation}

The total angular momentum $\mathbf{S}$ associated with the magnetic
moments is given by
\begin{equation}
\mathbf{S}=\gamma^{-1}\mathbf{M}=\sum_{i}\gamma^{-1}M_{i}\mathbf{e}_{i}.
\end{equation}
Therefore, angular momentum conservation is equivalent to the conservation
of the total magnetization $\mathbf{M}$. We have to consider two contributions,
\begin{equation}
\dot{\mathbf{M}}=\sum_{i}\left(\dot{M}_{i}\mathbf{e}_{i}+M_{i}\dot{\mathbf{e}}_{i}\right).\label{eq:M dot}
\end{equation}
The first term vanishes in general only if $\dot{M}_{i}=0$, i.e.
for constant moment lengths. For the second term, we find using the
equation of motion (\ref{eq:equation of motion}),
\begin{equation}
\sum_{i}M_{i}\dot{\mathbf{e}}_{i}=\sum_{i}\gamma M_{i}\mathbf{e}_{i}\times\mathbf{B}_{i}^{\text{eff}}.
\end{equation}
Assuming a Heisenberg-like effective magnetic field,
\begin{equation}
\mathbf{B}_{i}^{\text{eff}}=\frac{1}{M_{i}}\sum_{j}J_{ij}\mathbf{e}_{j},
\end{equation}
we obtain
\begin{equation}
\sum_{i}M_{i}\dot{\mathbf{e}}_{i}=\sum_{ij}\gamma J_{ij}\mathbf{e}_{i}\times\mathbf{e}_{j}=0,
\end{equation}
since by definition $J_{ij}=J_{ji}$. Therefore, only the second contribution
to $\dot{\mathbf{M}}$ in Eq.~(\ref{eq:M dot}) can be expected to vanish and we have to conclude
that adiabatic spin dynamics does not conserve the total angular momentum,
as angular momentum is exchanged with the bath if the moment lengths
are not constant. We note that $\sum_{i}M_{i}\dot{\mathbf{e}}_{i}=0$
does not imply $\sum_{i}\dot{\mathbf{e}}=0$ because the moment lengths
$M_{i}$ differ for each site in an arbitrary non-collinear state
without translational invariance.

\section{Effective Heisenberg exchange in non-collinear states\label{sec:Effective-Heisenberg-exchange}}

The energy curvature tensor describes the energy curvature with respect to pairwise rotations of the 
magnetic moments and is defined by
\begin{equation}
\mathcal{J}_{ij}^{\alpha\beta}=-\frac{\partial^{2}E}{\partial e_{j\beta}\partial e_{i\alpha}}=\frac{\partial (M_{i}B_{i\alpha}^{\text{eff}})}{\partial e_{j\beta}},
\end{equation}
where $E$ is the energy of the system. It should be noted that the derivatives with
respect to the moment directions $\mathbf{e}_{i}$ are taken with
the constraint of fixed length since they are unit vectors and can
only be rotated. We emphasize that this tensor $\mathcal{J}_{ij}^{\alpha\beta}$
is not equivalent to the exchange tensor ${J}_{ij}^{\alpha\beta}$ in a tensorial Heisenberg model with energy
\begin{equation}
E_{{J}}=-\frac{1}{2}\sum_{i\alpha,j\beta}{J}_{ij}^{\alpha\beta}{e}_{i\alpha}{e}_{j\beta}\label{eq:tensor Heisenberg}.
\end{equation}
The reason, as we further discuss below, is that for the calculation of the energy curvature of Eq.~(\ref{eq:tensor Heisenberg}) it is necessary to take the restriction to unit length into account, which implies ${\partial e_{j\alpha}}/{\partial e_{j\beta}}\neq\delta_{\alpha\beta}$, see Appendix~\ref{sec:Derivative-matrix}.  Therefore,
\begin{equation}
\mathcal{J}_{ij}^{\alpha\beta}\neq {J}_{ij}^{\alpha\beta},
\end{equation}
and we have to distinguish between the energy curvature tensor and the exchange tensor defined by Eq.~(\ref{eq:tensor Heisenberg}).

The curvature tensor $\mathcal{J}_{ij}^{\alpha\beta}$ can still be applied in a local spin Hamiltonian \cite{Streib2021} and the exchange tensor ${J}_{ij}^{\alpha\beta}$ can be extracted from the curvature tensor by a set of collinear configurations \cite{Udvardi2003}, but not from a single (non-collinear) configuration. Furthermore, we show below how in the case of isotropic exchange an effective exchange interaction can be derived from the curvature tensor in non-collinear configurations.

\subsection{Generalized exchange interaction\label{subsec:Generalized-exchange-interaction}}

We consider the following general two-spin isotropic exchange energy,
\begin{equation}
E=-\frac{1}{2}\sum_{ij}f_{ij}(\mathbf{e}_{i}\cdot\mathbf{e}_{j}),\label{eq:two-spin exchange}
\end{equation}
where $f_{ij}=f_{ji}$ (with $f_{ii}=0$) is a function of $\mathbf{e}_{i}\cdot\mathbf{e}_{j}$.
The effective magnetic field acting on spin $i$ is then
\begin{equation}
\mathbf{B}_{i}^{\text{eff}}=\frac{1}{M_{i}}\sum_{j}f_{ij}'(\mathbf{e}_{i}\cdot\mathbf{e}_{j})\mathbf{e}_{j},
\end{equation}
where $f_{ij}'$ denotes the derivative of $f_{ij}$ with respect to its argument $\mathbf{e}_{i}\cdot\mathbf{e}_{j}$.
We now define the effective exchange interaction by
\begin{equation}
J_{ij}(\theta_{ij})=f_{ij}'(\mathbf{e}_{i}\cdot\mathbf{e}_{j}),
\end{equation}
with
\begin{equation}
\mathbf{e}_{i}\cdot\mathbf{e}_{j}=\cos\theta_{ij}.
\end{equation}
Therefore, we write
\begin{equation}
\mathbf{B}_{i}^{\text{eff}}=\frac{1}{M_{i}}\sum_{j}J_{ij}(\theta_{ij})\mathbf{e}_{j}.\label{eq:effective field}
\end{equation}
To take the fixed length of the unit vectors into account, we have
to consider the perpendicular part of this effective field,
\begin{equation}
\mathbf{B}_{i,\perp}^{\text{eff}}=\mathbf{B}_{i}^{\text{eff}}-\mathbf{e}_{i}\left(\mathbf{B}_{i}^{\text{eff}}\cdot\mathbf{e}_{i}\right).\label{eq:perp effective field}
\end{equation}
We note that
\begin{equation}
J_{ij}'(\theta_{ij})=\frac{\partial J_{ij}(\theta_{ij})}{\partial\theta_{ij}}=-f_{ij}''(\cos\theta_{ij})\sin\theta_{ij},\label{eq:J'}
\end{equation}
implying that $J'_{ij}(\theta_{ij})$ is expected to vanish for $\theta_{ij}=0$.

\subsection{Determining the effective exchange} \label{sec:effective exchange}

We can now calculate the energy curvature tensor $\mathcal{J}_{ij}^{\alpha\beta}$ from
the effective field (\ref{eq:perp effective field}). We assume a reference
coordinate system where $\mathbf{e}_{i}=\hat{\mathbf{z}}$ and $\mathbf{e}_{j}$
is rotated by an angle $\theta_{ij}$ in the $xz$ plane. Here, it
is crucial that we take the derivative of the
effective field, Eq.~(\ref{eq:perp effective field}), with respect
to unit vectors, see Appendix~\ref{sec:Derivative-matrix}. We obtain
for $i\neq j$,
\begin{align}
\mathcal{J}_{ij}^{xx} & =J_{ij}(\theta_{ij})\cos^{2}\theta_{ij}+\sin\theta_{ij}\cos\theta_{ij}J'_{ij}(\theta_{ij}),\label{eq:Jxx}\\
\mathcal{J}_{ij}^{yy} & =J_{ij}(\theta_{ij}),\label{eq:Jyy}\\
\mathcal{J}_{ij}^{xz} & =-J_{ij}(\theta_{ij})\sin\theta_{ij}\cos\theta_{ij}-\sin^{2}\theta_{ij}J'_{ij}(\theta_{ij}),\label{eq:Jxz}\\
\mathcal{J}_{ij}^{xy} & =\mathcal{J}_{ij}^{yx}=\mathcal{J}_{ij}^{yz}=\mathcal{J}_{ij}^{zx}=\mathcal{J}_{ij}^{zy}=\mathcal{J}_{ij}^{zz}=0.
\end{align}
The effective exchange $J_{ij}(\theta_{ij})$ can therefore be obtained
from $\mathcal{J}_{ij}^{yy}$ in the coordinate system as specified above. The
$yy$ component of $\mathcal{J}_{ij}^{\alpha\beta}$ in this reference coordinate
system corresponds to variations of the moment directions perpendicular
to the plane that they span. In a general case, we can always rotate
the tensor $\mathcal{J}_{ij}^{\alpha\beta}$ from a global coordinate system
to this specific reference coordinate system for each pair $(i,j)$.
From these results, we see that even for an ideal Heisenberg exchange
with $J_{ij}'=0$, we have $\mathcal{J}_{ij}^{xx}\neq \mathcal{J}_{ij}^{yy}$ and $\mathcal{J}_{ij}^{xz}\neq0$
in a non-collinear state with $\theta_{ij}\neq0$, which could
be mistaken for non-Heisenberg behavior. For the contribution of a Dzyaloshinskii–Moriya interaction (DMI) \cite{Dzyaloshinsky1958,Moriya1960}
to the energy curvature tensor, see Appendix~\ref{sec:DMI}.

\begin{figure}
\begin{centering}
\includegraphics{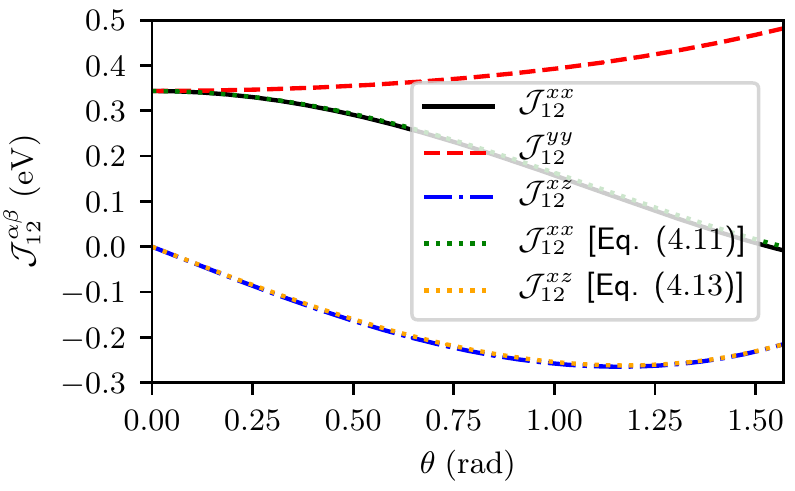}
\par\end{centering}
\caption{Comparison of the energy curvature components $\mathcal{J}_{12}^{xx}$ and $\mathcal{J}_{12}^{xz}$
of an Fe dimer \cite{Streib2021} with results obtained from
Eqs.~(\ref{eq:Jxx}-\ref{eq:Jxz}) by setting $J_{12}=\mathcal{J}_{12}^{yy}$.\label{fig:J dimer}}

\end{figure}

\begin{figure}
\begin{centering}
\includegraphics{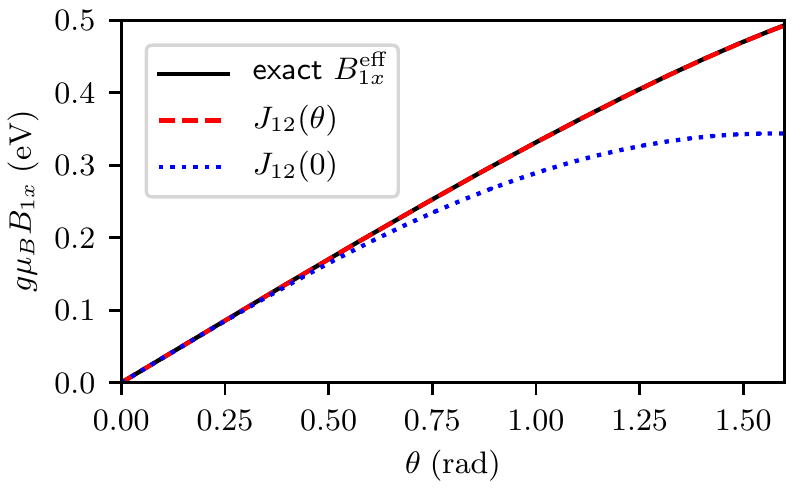}
\par\end{centering}
\caption{Comparison of the exact effective magnetic field $B_{1x}^{\text{eff}}$,
acting on the first moment of an Fe dimer when the second moment
is rotated by $\theta$ \cite{Streib2021}, with the effective field
obtained from the effective exchange $J_{12}(\theta)$ and from the Heisenberg
model with constant exchange $J_{12}(0)$. \label{fig:Dimer B}}

\end{figure}

We check the consistency of Eqs.~(\ref{eq:Jxx}-\ref{eq:Jxz}) by
applying them to results previously obtained for an Fe dimer without
spin-orbit coupling \cite{Streib2021}, where the second moment is
rotated by an angle $\theta$. We obtain $J_{12}$ and $J_{12}'$
from $\mathcal{J}_{12}^{yy}$ and plug these quantities into Eqs.~(\ref{eq:Jxx})
and (\ref{eq:Jxz}), which are shown together with the previously
calculated exchange parameters in Fig.~\ref{fig:J dimer}. 
The agreement is excellent and the small deviations are due to approximations made
in the calculation of the curvature tensor, see Ref.~\cite{Streib2021}.
For a
dimer without spin-orbit coupling, the two-spin exchange interaction (\ref{eq:two-spin exchange})
is expected to be exact, which we confirm in Fig.~\ref{fig:Dimer B}
by comparing the exact effective field obtained from the constraining
field with the effective field (\ref{eq:perp effective field}) obtained
from the exchange $J_{12}=\mathcal{J}_{12}^{yy}$ via Eq.~(\ref{eq:effective field}).
We note that for systems consisting of more than two magnetic moments, 
multi-spin interactions \cite{Tanaka1977,MacDonald1988,Singer2011,Hoffmann2020,Brinker2020,Mankovsky2020}
can also contribute to the effective field and 
the two-spin exchange interaction cannot be expected to provide the exact effective field.

When calculating the energy curvature tensor $\mathcal{J}_{ij}^{\alpha\beta}$
from the adiabatic energy (\ref{eq:adiabatic energy}), which produces an effective field that is perpendicular
to the moment directions due to the minimization with respect to the moment lengths, it is not necessary to explicitly take the
fixed unit length of the unit vectors $\mathbf{e}_{i}$ and $\mathbf{e}_{j}$
into account for $i\neq j$, i.e., it is allowed to just take the
Cartesian derivatives with respect to the components $e_{i\alpha}$
and $e_{j\beta}$. This follows from the fact that the restriction
of unit length is equivalent to setting the component of the gradient
$\boldsymbol{\nabla}_{\mathbf{e}_{i}}$ parallel to $\mathbf{e}_{i}$
to zero, see Eq.~(\ref{eq:perp effective field}). If the gradient
of the energy has no such parallel component, it is therefore not necessary
to apply this restriction. For the second derivative $\partial/\partial e_{j\beta}$,
we have to project out the component parallel to $\mathbf{e}_{j}$,
\begin{align}
\left.\boldsymbol{\nabla}_{\mathbf{e}_{j}}\frac{\partial E}{\partial e_{i\alpha}}\right|_{\perp} & =\boldsymbol{\nabla}_{\mathbf{e}_{j}}\frac{\partial E}{\text{\ensuremath{\partial}}e_{i\alpha}}-\mathbf{e}_{j}\left(\mathbf{e}_{j}\cdot\boldsymbol{\nabla}_{\mathbf{e}_{j}}\frac{\partial E}{\partial e_{i\alpha}}\right).
\end{align}
For $i\neq j$, we can pull the derivative with respect to $e_{i\alpha}$
in front and use that $\mathbf{e}_{j}\cdot\boldsymbol{\nabla}_{\mathbf{e}_{j}}E=0$,

\begin{align}
\left.\boldsymbol{\nabla}_{\mathbf{e}_{j}}\frac{\partial E}{\partial e_{i\alpha}}\right|_{\perp} & \overset{i\ne j}{=}\boldsymbol{\nabla}_{\mathbf{e}_{j}}\frac{\partial E}{\partial e_{i\alpha}}-\frac{\partial}{\partial e_{i\alpha}}\mathbf{e}_{j}\left(\mathbf{e}_{j}\cdot\boldsymbol{\nabla}_{\mathbf{e}_{j}}E\right)\nonumber \\
 & \overset{i\ne j}{=}\boldsymbol{\nabla}_{\mathbf{e}_{j}}\frac{\partial E}{\partial e_{i\alpha}}.
\end{align}
Therefore, the projection is then not required for $i\neq j$. This
consideration does not apply to the tensorial Heisenberg model (\ref{eq:tensor Heisenberg}) and the exchange
energy (\ref{eq:two-spin exchange}) above since the corresponding
effective fields have a parallel component,
which requires to take the restriction to unit length explicitly
into account.

In summary, the effective exchange interaction $J_{ij}$ can be obtained
from the following procedure:
\begin{enumerate}
\item Calculate the energy curvature tensor $\mathcal{J}_{ij}^{\alpha\beta}$.
\item For each pair $(i,j)$, rotate $\mathcal{J}_{ij}^{\alpha\beta}$ to the coordinate
system where $\mathbf{e}_{i}=\hat{\mathbf{z}}$ and $\mathbf{e}_{j}$
lies in the $xz$ plane.
\item Set $J_{ij}=\mathcal{J}_{ij}^{yy}$.
\end{enumerate}

\section{Tight-binding model\label{sec:Tight-binding-model}}

The tight-binding electronic structure model employed here is implemented in the software package \textit{Cahmd} \cite{CAHMD} and is based on a Hamiltonian that consists of a hopping
term $\hat{\mathcal{H}}_{0}$, a local charge neutrality term $\hat{\mathcal{H}}_{\mathrm{lcn}}$,
and a Stoner term $\hat{\mathcal{H}}_{\text{St}}$ \cite{Aute2006,Rossen2019,Cardias2021a},
\begin{equation}
\hat{\mathcal{H}}_{\mathrm{tb}}=\hat{\mathcal{H}}_{0}+\hat{\mathcal{H}}_{\mathrm{lcn}}+\hat{\mathcal{H}}_{\text{St}}.\label{eq:H_tb}
\end{equation}
The hopping term is in second-quantization given by 
\begin{equation}
\hat{\mathcal{H}}_{0}=\sum_{i\ell,j\ell',\sigma}t_{i\ell,j\ell'}\hat{c}_{i\ell\sigma}^{\dagger}\hat{c}_{j\ell'\sigma},\label{eq:hopping}
\end{equation}
which describes the hopping of an electron from state $j\ell'\sigma$
to $i\ell\sigma$ with hopping amplitude $t_{i\ell,j\ell'}$ and creation
and annihilation operators $\hat{c}_{i\ell\sigma}^{\dagger}$ and
$\hat{c}_{j\ell'\sigma}$. The index $i\ell\sigma$ indicates the
lattice site, orbit, and spin, respectively. We use a Slater-Koster parameterization \cite{Slater1954} of the hopping parameters
in a non-orthognal basis which is based on Ref.~\cite{Mehl1996}. It has been shown that such a tight-binding model provides a numerically efficient and valid description of transition metal elements and alloys \cite{Barreteau2016} both for collinear and non-collinear magnetic configurations, based on comparisons with DFT calculations \cite{Soulairol2016}.

The local charge neutrality term is given by
\begin{align}
\hat{\mathcal{H}}_{\text{lcn}} & =\frac{1}{2}U_{\text{lcn}}\sum_{i}\left(\hat{n}_{i}-n_{i}^{0}\right)\left(\hat{n}_{i}-n_{i}^{0}\right)\nonumber\\
 & \approx U_{\text{lcn}}\sum_{i}\hat{n}_{i}\left(n_{i}-n_{i}^{0}\right)+E_{\text{lcn}}^{\text{dc}},
\end{align}
where we apply a mean-field approximation and the double-counting
contribution is 
\begin{equation}
E_{\text{lcn}}^{\text{dc}}=-\frac{1}{2}U_{\text{lcn}}\sum_{i}\left[n_{i}^{2}-(n_{i}^{0})^{2}\right].
\end{equation}
Here,
\begin{equation}
\hat{n}_{i}=\sum_{\ell\sigma}\hat{c}_{i\ell\sigma}^{\dagger}\hat{c}_{i\ell\sigma}
\end{equation}
counts the number of electrons at site $i$ and $n_{i}^{0}$ is the
prescribed number of electrons per site with $n_{i}=\langle\hat{n}_{i}\rangle$.
The local charge neutrality term is similar to a Coulomb interaction
and favors a charge-neutral state, which is exactly enforced in the limit $U_\mathrm{lcn}\to\infty$. In practice, $U_\mathrm{lcn}$ should be a large positive quantity \cite{Barreteau2016,Soulairol2016} and we use here $U_{\text{lcn}}=5\;\text{eV}$ \cite{Schena2010,Rossen2019}. If all atoms are geometrically and chemically equivalent, the local charge neutrality term is only required for non-collinear magnetic configurations that break the equivalence. Finally,
we describe the spin-splitting with a Stoner contribution $\hat{\mathcal{H}}_{\text{St}}$
to the Hamiltonian, which we discuss in Sec.~\ref{subsec:Stoner-model},
and we explain the consequences for the calculation of the energy curvature
tensor in Sec.~\ref{subsec:Calculation-of-J}. We note that we do not include spin-orbit coupling in this work.

\subsection{Stoner model\label{subsec:Stoner-model}}

For the construction of the Stoner contribution $\hat{\mathcal{H}}_{\text{St}}$
to the Hamiltonian, it is crucial to ensure that 
\begin{equation}
\left\langle \boldsymbol{\nabla}_{\mathbf{e}_{i}}\hat{\mathcal{H}}_{\text{St}}\right\rangle =0,
\end{equation}
such that the energy is conserved. In some previous works \cite{Streib2020,Streib2021},
a Stoner term of the following form was used,
\begin{equation}
\hat{\mathcal{H}}_{\text{St}}^{\text{old}}=-\frac{1}{2}\sum_{i\ell\ell'}I_{\ell\ell'}M_{i\ell}\mathbf{e}_{i}\cdot\hat{\mathbf{M}}_{i\ell'},\label{eq:old Stoner}
\end{equation}
where this requirement is not fulfilled and it is not possible to
obtain a conserved total energy. Since we consider the conservation
of energy as a necessary requirement for our spin dynamics simulations,
we use instead the following Stoner term within a mean-field
approximation \cite{Aute2006, Rossen2019,Cardias2021a},
\begin{align}
\hat{\mathcal{H}}_{\text{St}} & =-\frac{1}{4}\sum_{i\ell\ell'}I_{\ell\ell'}\hat{\mathbf{M}}_{i\ell}\cdot\hat{\mathbf{M}}_{i\ell'}\nonumber \\
 & \approx-\frac{1}{2}\sum_{i\ell\ell'}I_{\ell\ell'}\mathbf{M}_{i\ell}\cdot\hat{\mathbf{M}}_{i\ell'}+E_{\text{St}}^{\text{dc}},\label{eq:new Stoner}
\end{align}
where the double-counting term is given by

\begin{equation}
E_{\text{St}}^{\text{dc}}=\frac{1}{4}\sum_{i\ell\ell'}I_{\ell\ell'}\mathbf{M}_{i\ell}\cdot\mathbf{M}_{i\ell'}.
\end{equation}
The inclusion of the energy contribution $E_{\text{St}}^{\text{dc}}$
is required to fulfill the condition $\langle\boldsymbol{\nabla}_{\mathbf{e}_{i}}\hat{\mathcal{H}}_{\text{St}}\rangle=0$.

\subsection{Calculation of the energy curvature tensor\label{subsec:Calculation-of-J}}

The procedure in Sec.~\ref{sec:effective exchange} to obtain the effective exchange interaction $J_{ij}$ requires the energy curvature tensor $\mathcal{J}_{ij}^{\alpha\beta}$.
For the calculation of $\mathcal{J}_{ij}^{\alpha\beta}$ 
within the above specified tight-binding description, we have to
evaluate the gradient of the Stoner term, $\boldsymbol{\nabla}_{\mathbf{e}_{j}}\hat{\mathcal{H}}_{\text{St}}$, see Ref.~\cite{Streib2021} for details on the formalism.
While this is straight-forward for the previous implementation, Eq\@.~(\ref{eq:old Stoner}),
in the case of our current implementation, Eq.~(\ref{eq:new Stoner}),
this involves the calculation of

\begin{equation}
\frac{\partial}{\partial e_{j\beta}}M_{i\ell\alpha}\approx M_{i\ell}\frac{\partial e_{i\alpha}}{\partial e_{j\beta}}=M_{i\ell}\delta_{ij}\delta_{\alpha\beta},\label{eq:orbital approximation}
\end{equation}
which we estimate by assuming that all orbital contributions point
along the same direction, $\mathbf{M}_{i\ell}=M_{i\ell}\mathbf{e}_{i}$. Taking here the full derivative, ${\partial e_{i\alpha}}/{\partial e_{i\beta}}=\delta_{\alpha\beta}$, results in unphysical contributions to the energy gradient tensor which can be projected out and do not affect the physically relevant contributions \cite{Streib2021}.
We observe that the approximation in Eq.~(\ref{eq:orbital approximation}) does not result in accurate
results when comparing the effective exchange $J_{ij}^\text{c}$ calculated from the energy curvature tensor given by Eq.~(3.23) in Ref.~\cite{Streib2021}
with the exchange obtained from the exact energy curvature tensor calculated by numerical differentiation of the
constraining field. However, it turns out that an alternative result $J_{ij}^\text{sc}$ based on the
curvature of the band energy, Eq.~(3.28) in Ref.~\cite{Streib2021}, which is
analogous to Eq.~(11) in Ref.~\cite{Bruno2003}, gives
more accurate results within the approximation (\ref{eq:orbital approximation}). For example, for
the Fe chain discussed in Sec.~\ref{subsec:Iron-spin-chain}, we
obtain in the ferromagnetic ground state
\begin{align}
J_{i,i+1}^{\text{exact}} & =0.1298\;\text{eV},\\
J_{i,i+1}^{\text{c}} & =0.1057\;\text{eV},\\
J_{i,i+1}^{\text{sc}} & =0.1308\;\text{eV}.
\end{align}
Furthermore, in the case of an Fe dimer, we have also confirmed the
reliability of $J_{ij}^{\text{sc}}$ in non-collinear configurations.
Therefore, we will use in the following $J_{ij}^{\text{sc}}$ to provide
an estimate of the exchange interaction $J_{ij}$. We note that more rigorous
results could be achieved by calculating $\frac{\partial}{\partial e_{j\beta}}M_{i\ell\alpha}$
self-consistently together with the energy curvature tensor, which is beyond
our current implementation.

\section{Dynamics and effective exchange of spin-disordered states\label{sec:Dynamics}}

\subsection{Fe, Co and Ni dimers
}
\begin{figure}
\begin{centering}
\includegraphics[width=1\linewidth]{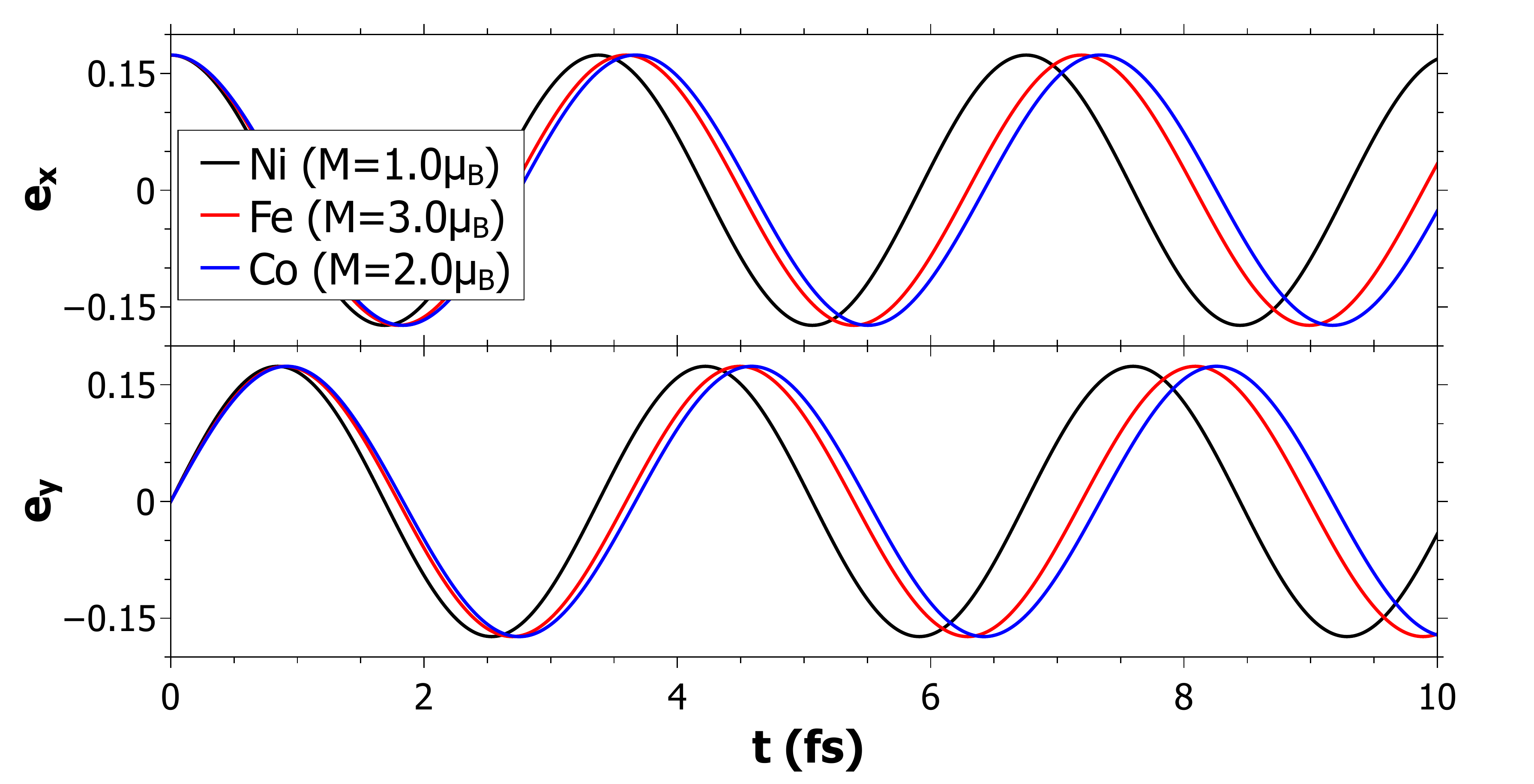}
\par\end{centering}
\caption{Time evolution of the magnetic moment components $e_x$ and $e_y$ for Ni, Fe and Co dimers (black, red and blue, respectively). Calculations are performed without damping. Note that since the calculation was performed without damping, the $z$ component of the magnetic moment is kept constant. The curves are for one atom. The other atom has an inverted dynamics due to the symmetry of the system.\label{fig:sd1}}

\end{figure}

To test the tight-binding implementation of the \textit{Cahmd} code \cite{CAHMD}, we have considered dimers of Fe, Co and Ni with a lattice constant of $\unit[2.0]{\textup{\AA}}$ and studied their dynamics. We took initially an angle of $\theta=20^\circ$, with $\theta$ being the angle between the magnetic moments, and performed the calculations with and without damping. The results without damping showed a good agreement with previous work~\cite{Cardias2021a}, with the results shown in Fig.~\ref{fig:sd1}.

\begin{figure}
\begin{centering}
\includegraphics[width=0.95\linewidth]{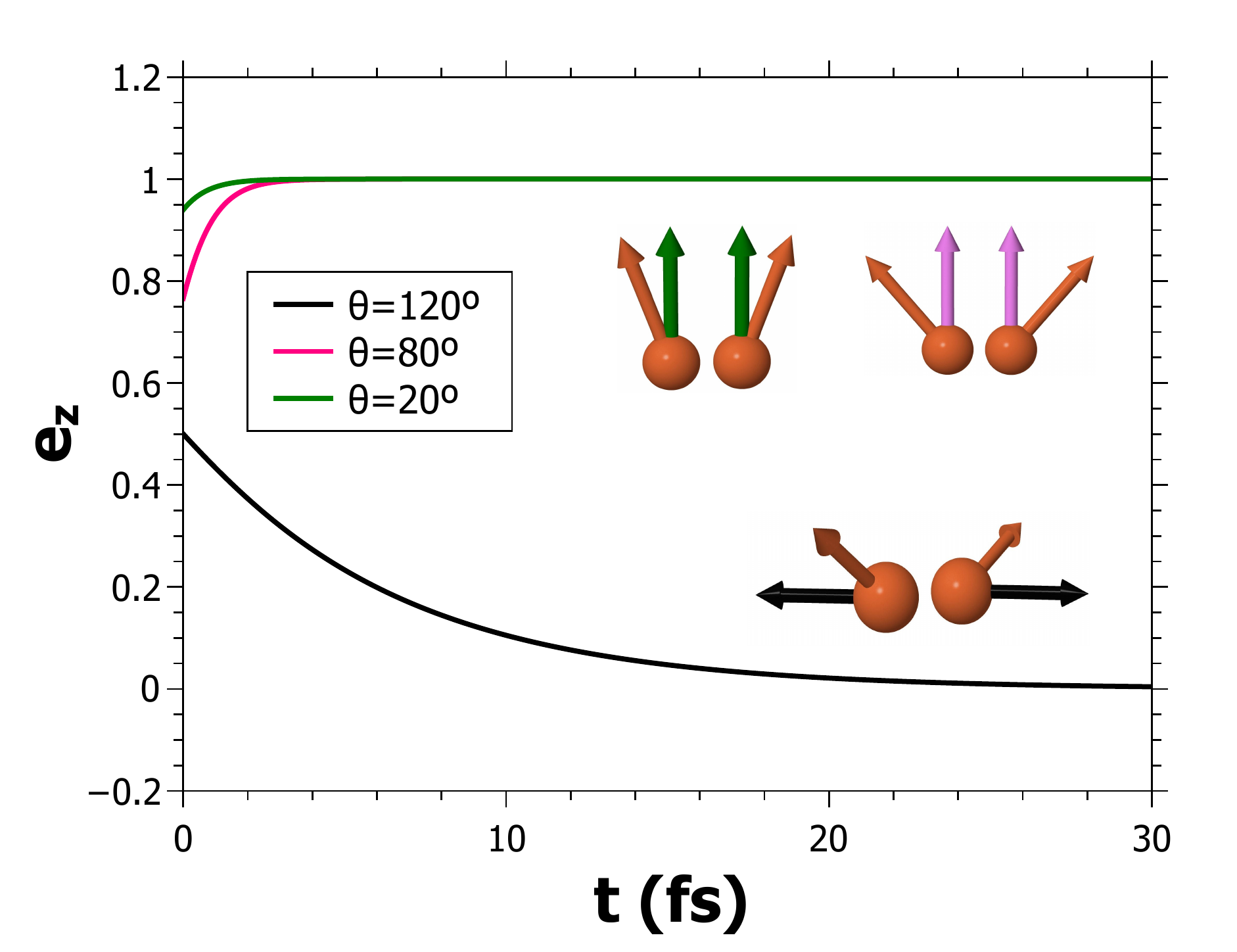}
\par\end{centering}
\caption{Time evolution of the $z$ component of the Fe dimer moment directions. Calculations were performed with the damping $\alpha=0.5$. The orange color of the atoms stands for the initial configuration whereas the different colored arrows represent the final configurations concerning their respective color in the legend. \label{fig:sd2}}

\end{figure}

\begin{figure}
\begin{centering}
\includegraphics[width=1\linewidth]{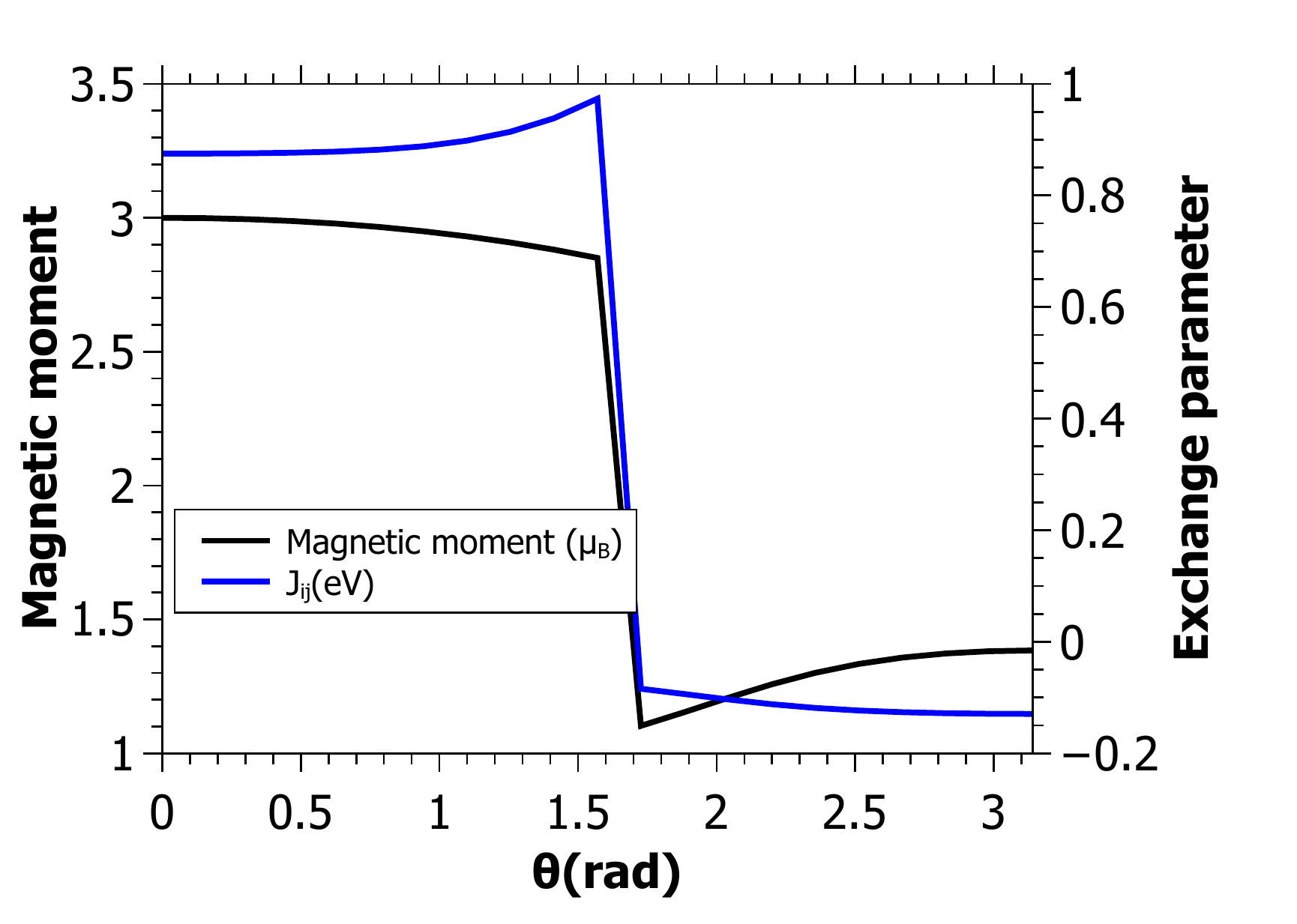}
\par\end{centering}
\caption{The calculated exchange parameter $J$ for different angles when varying a single magnetic moment in the Fe dimer (right axis) and the calculated magnetic moment for each one of these magnetic configurations (left axis).\label{fig:j_x_spin}}

\end{figure}

For the systems with damping, we focused on the case of Fe. Here, we used a large damping $\alpha=0.5$ to let the system quickly relax to the ground state starting from different magnetic configurations (e.g., $\theta=20^\circ$, $\theta=80^\circ$ and $\theta=120^\circ$), as shown in Fig.~\ref{fig:sd2}. In the cases of $\theta=20^\circ$ and $\theta=80^\circ$, the dimer relaxed to a ferromagnetic (FM) state. For the angle $\theta=120^\circ$, the Fe dimer relaxed to the anti-ferromagnetic (AFM) configuration. This is in agreement with the sign of the exchange coupling parameter calculated for each starting configuration, see Fig.~\ref{fig:j_x_spin}, as well as with the results shown in Ref.~\cite{Cardias2021a} where a stable ground state can be found around $\theta=180^\circ$.

Previous works have reported the nature of the AFM behavior of a few 3d metals by analyzing the different orbital-orbital contribution to the total exchange parameters in a crystal~\cite{Kvashnin2016,Cardias2017}, in particular Fe having a strong AF contribution coming from the T$_{2g}$-T$_{2g}$ orbitals, therefore it is not surprising that the Fe dimer has a negative exchange coupling for a given magnetic configuration. In Fig.~\ref{fig:j_x_spin} one can see the dependence of the exchange parameter $J_{ij}$ as a function of the angle between the magnetic moments. Note that there is a discontinuity around 70 degrees and that is due to the abrupt change to the magnetic moment from approximately $3.0\;\mu_{B}$ for small angles to $1.22\;\mu_{B}$ at $\theta = 120^{\circ}$, revealing the strong non-Heisenberg behavior for this system.

Additionally, we also performed calculations for Fe and Co dimers using an orthogonal tight-binding basis, initially applied in Ref. \cite{Thonig2014}. The results (data not shown) show that the electronic structure is very similar to the one obtained using the non-orthogonal basis. This leads to similar exchange parameters and therefore  similar dynamics of the magnetic moments compared to the ones presented in this section.

\subsection{Fe chain\label{subsec:Iron-spin-chain}}

\begin{figure}
\begin{centering}
\includegraphics{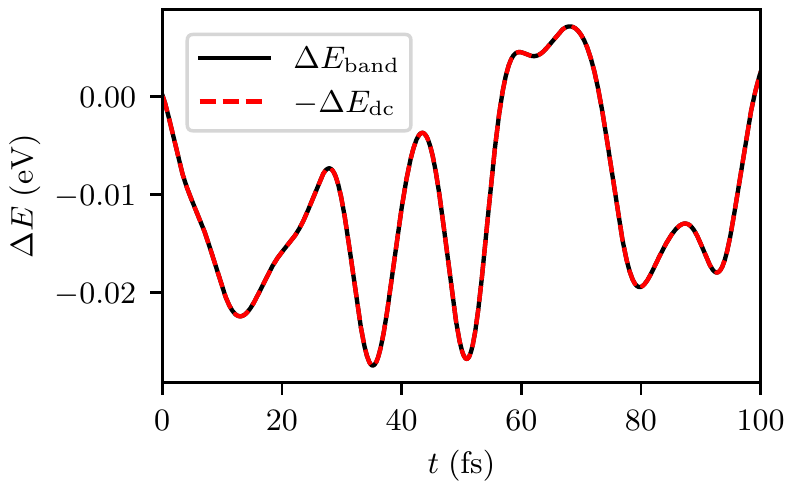}
\par\end{centering}
\caption{Time evolution of the change of the band energy ($\Delta E_{\text{band}}$) and double
counting ($\Delta E_{\text{dc}}$) contributions to the total energy $E=E_{\text{band}}+E_{\text{dc}}$
of a 10 atom Fe spin chain in a non-collinear configuration without
damping, demonstrating energy conservation.\label{fig:energy conservation}}
\end{figure}

\begin{figure}
\begin{centering}
\includegraphics{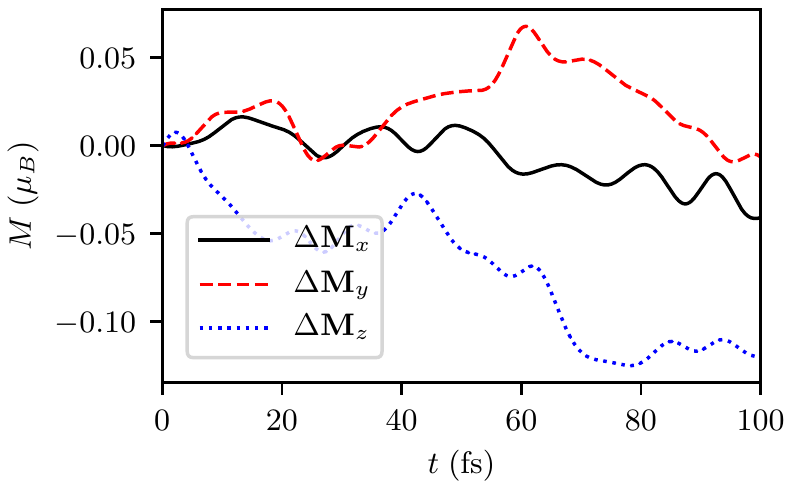}
\par\end{centering}
\caption{Time evolution of the change $\Delta\mathbf{M}$ of the components of the total magnetization $\mathbf{M}$
of a 10 atom Fe spin chain from a non-collinear configuration without
damping.\label{fig:angular momentum}}

\end{figure}

\begin{figure}
\begin{centering}
\includegraphics{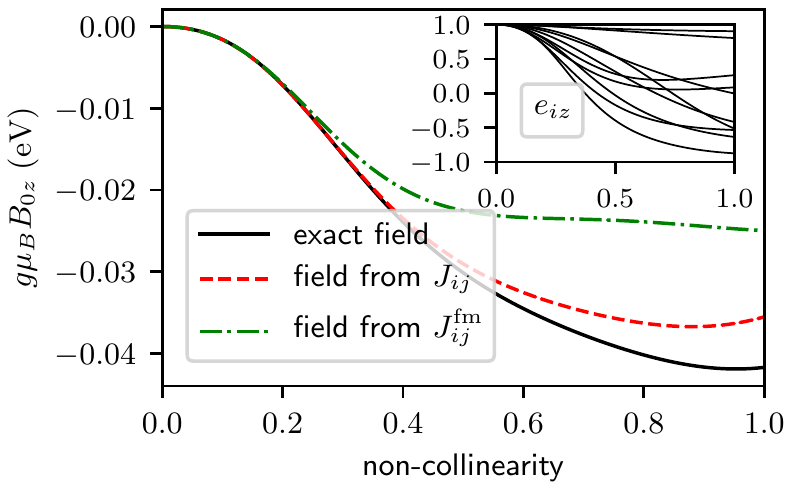}
\par\end{centering}
\caption{Comparison of the $z$ component of the exact effective field acting on the magnetic moment at site $i=0$ in a periodic 10 atom Fe spin chain with effective fields obtained from the effective exchange interaction $J_{ij}$ and its ferromagnetic ground-state value $J_{ij}^\mathrm{fm}$. The non-collinearity parameter describes a continuous transformation from a collinear state to a random non-collinear state, as illustrated in the inset by the $z$ components of the moment vectors $\mathbf{e}_i$.\label{fig:disorder}}
\end{figure}

We consider next the dynamics of an Fe spin chain with a lattice
constant of $2.486\;\mathring{\mathrm{A}}$, which consists of $10$
atoms with periodic boundary conditions. We first check the analytical
results on the conservation of energy and non-conservation of angular
momentum by running a short $100\;\mathrm{fs}$ simulation without damping, starting from a
randomly oriented spin configuration given by the maximally non-collinear configuration indicated in the inset of Fig.~\ref{fig:disorder}. Figure \ref{fig:energy conservation}
shows that the change of the double counting contribution $E_{\text{dc}}$ compensates the change of
the band energy $E_{\text{band}}$ such that the total energy is conserved within
a numerical accuracy that depends on the chosen time step length. Here, with time steps of $0.1\;\mathrm{fs}$, the fluctuations are of the order of several $\mu\mathrm{eV}$. The non-conservation of angular momentum can be seen in Fig.~\ref{fig:angular momentum},
which shows the components of the total magnetization vector $\mathbf{M}$. We note that if we artificially constrain the moment lengths to a fixed value, angular momentum is conserved in our simulations as expected.

\begin{figure}
\begin{centering}
\includegraphics{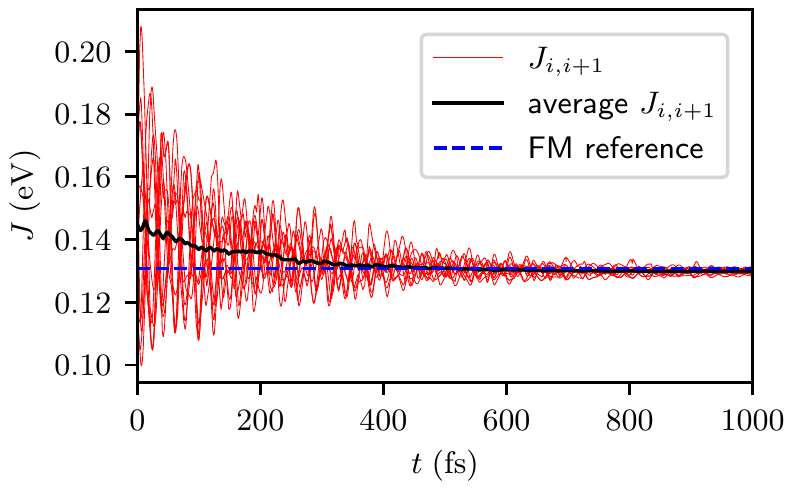}
\par\end{centering}
\caption{Relaxation of the nearest-neighbor exchange $J_{i,i+1}$ of a 10 atom
Fe spin chain from a non-collinear configuration with damping $\alpha=0.01$.\label{fig:J damping}}
\end{figure}

\begin{figure}
\begin{centering}
\includegraphics{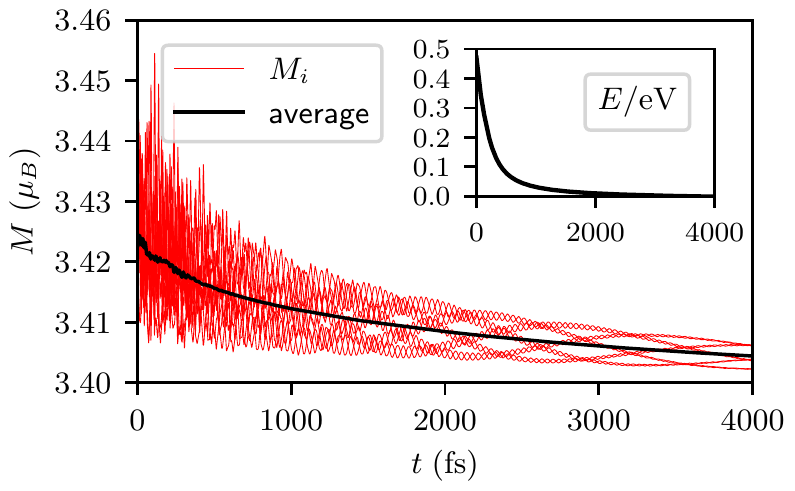}
\par\end{centering}
\caption{Relaxation of the magnetic moment lengths $M_{i}$ of a 10 atom
Fe spin chain from a non-collinear configuration with damping $\alpha=0.01$.
Inset: total energy $E$.\label{fig:M damping}}
\end{figure}

Next, we switch on a damping of $\alpha=0.01$ and let the spin chain
relax. We track the nearest neighbor exchange interactions, magnetic
moments, and energy, which are shown in Figs.~\ref{fig:J damping}
and \ref{fig:M damping}, respectively. While the magnetic moments
show only small flucuations of the order of 1\%, the nearest neighbor
exchange can vary by more than 50\% and the average is increased by
about 10\% compared to the ferromagnetic ground state. Both energy
and exchange are mostly relaxed after a simulation time of $4000\;\text{fs}$,
but the individual magnetic moments still oscillate slowly and have
not reached their ground-state value of $3.4\;\mu_{B}$.

The Fe spin chain also allows to demonstrate the limitations of the effective
exchange interaction in strongly non-collinear states due to multi-spin interactions that are not
included in the two-spin exchange energy (\ref{eq:two-spin exchange}). In Fig.~\ref{fig:disorder},
we compare the exact effective field obtained from the constraining field with the effective field
obtained from the exchange interaction $J_{ij}$ via Eqs.~(\ref{eq:effective field}) and (\ref{eq:perp effective field}) under a continuous transformation of the moment directions from the ferromagnetic state to a random non-collinear state.
Close to the collinear configuration, the agreement is nearly perfect both with the field obtained from $J_{ij}$ updated for each configuration and from constant $J_{ij}^\mathrm{fm}$ obtained from the ferromagnetic ground state.
We find an improved agreement with the updated $J_{ij}$ over $J_{ij}^\mathrm{fm}$ for increasingly non-collinear states, but in strongly non-collinear states the effective exchange $J_{ij}$ is insufficient to obtain the correct effective field.
This is different from the dimer case where the effective exchange gives the exact effective field. Therefore, we attribute this failure of the isotropic two-spin exchange to multi-spin interactions that are not included in an effective pair exchange formalism, since the isotropy could only be broken by spin-orbit coupling, which is not included in the tight-binding Hamiltonian (\ref{eq:H_tb}) considered here.

\subsection{Co chain\label{subsec:Cobalt-spin-chain}}
\begin{figure}
\begin{centering}
\includegraphics{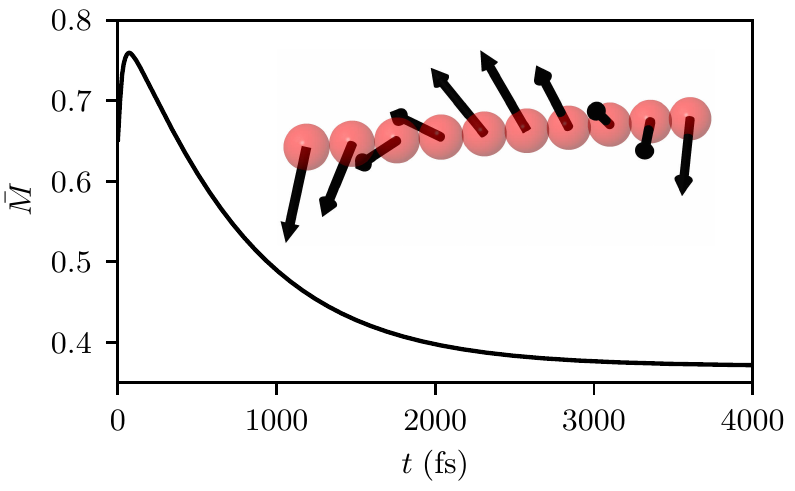}
\par\end{centering}
\caption{Time evolution of the normalized average magnetic moment $\bar{M}$ for $10$ atom chain of Co ($\bar{M}=1$ is the fully ferromagnetic state). The inset shows the spin configuration (black arrows) after equilibration.  \label{fig:Co:MvsTime}}
\end{figure}

We also investigated a chain of $10$ Co atoms along $x$ direction with a lattice constant of $\unit[2.50]{\textup{\AA}}$ in Fig.~\ref{fig:Co:MvsTime}. 
The Gilbert damping is set to $0.05$ in order to guarantee fast relaxation from the initial random state, which is the same as for the Fe chain, see inset of Fig.~\ref{fig:disorder}. The magnetic moment length of each spin is $M=\unit[2.35]{\mu_B}$ and approximately constant during the relaxation process with  a maximal deviation of $\Delta M=\unit[0.04]{\mu_B}$. 

\begin{figure}
\begin{centering}
\includegraphics{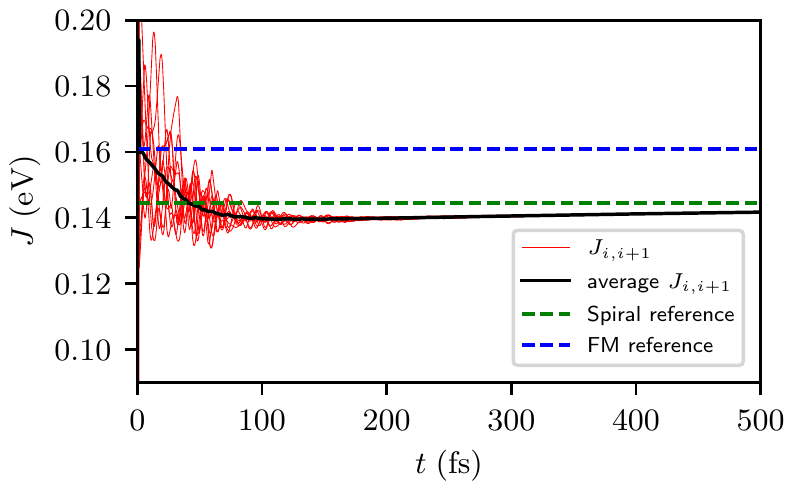}
\par\end{centering}
\caption{Relaxation of the nearest-neighbor exchange $J_{i,i+1}$ of a $10$ atom
Co spin chain from a non-collinear configuration with damping $\alpha=0.05$.\label{fig:Co:J}}
\end{figure}

Surprisingly, the normalized magnetization relaxes not to a ferromagnetic but to a spiral state (see inset in Fig.~\ref{fig:Co:MvsTime}), which does not depend on a particular choice of the damping parameter $\alpha$. From the total energy (per atom) of both the ferromagnetic state $E=\unit[3.558]{eV}$ and the spiral state $E=\unit[3.561]{eV}$ (so $\Delta E=\unit[3]{meV}=0.021J_{i,i+1}$), it is revealed that the spiral state is only metastable and not the ground state. Since the dynamics is driven by the exact effective field $\mathbf{B}_{i}^{\text{eff}}=-\mathbf{B}_{i}^{\text{con}}$, it is unclear which spin-spin exchange mechanism stabilizes the spiral state. Simulating the time-resolved Heisenberg exchange (Fig.~\ref{fig:Co:J}) shows a strong ferromagnetic coupling between the nearest-neighbor magnetic moments ($J_{i,i+1}>0$); the second nearest-neighbor couplings are typically two-orders of magnitude smaller and antiferromagnetic ($J_{i,i+2}<0$). Spiral states in $3d$ transition metal chains have been previously reported \cite{Tung2011,Toews2012}. No spiral ground state was found for Co, but in the case of Fe a spiral state was obtained for lattice constants below the bulk value of $2.486\;\mathring{\mathrm{A}}$ and a ferromagnetic ground state for the bulk value \cite{Toews2012}, which is consistent with the result for the Fe chain above.

\begin{figure}
\begin{centering}
\includegraphics{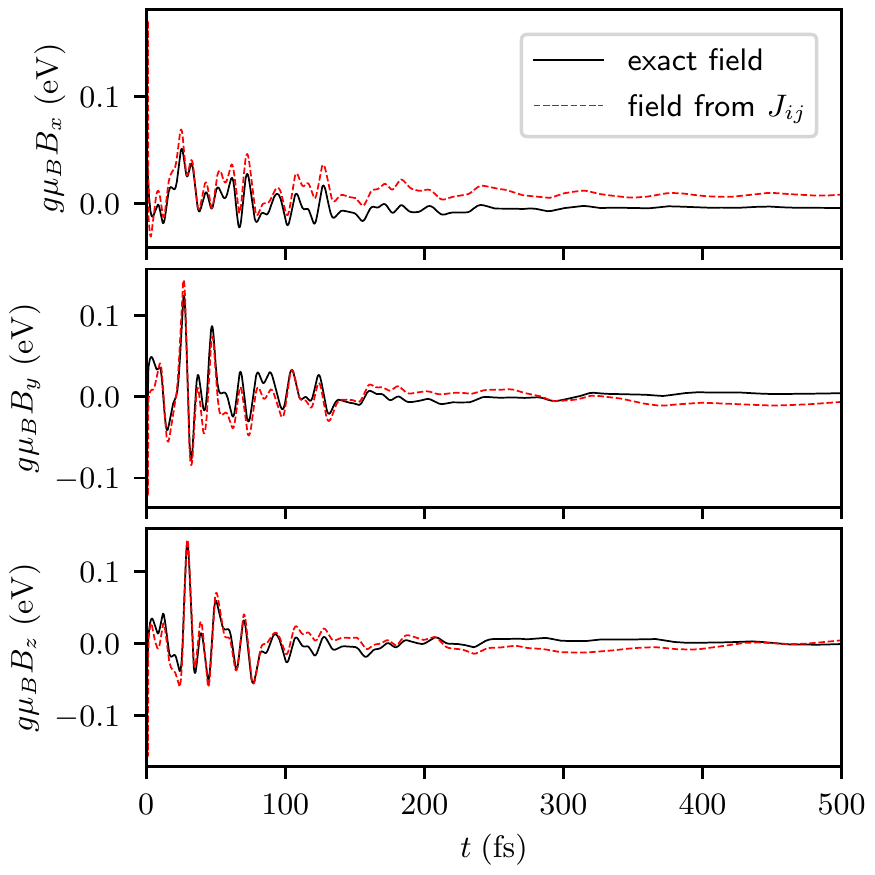}
\par\end{centering}
\caption{Dynamics of the exact effective field (black solid lines) and the effective field related to a Heisenberg spin-Hamiltonian with dynamical determined $J$'s (red dotted lines) for one atom in a chain of $10$ Co atoms. The different panels show the different Cartesian components of the fields.\label{fig:Co:B}}
\end{figure}

Classical atomistic magnetization dynamics from a Heisenberg spin-Hamiltonian shows opposite to the tight-binding dynamics a relaxation to the ferromagnetic state (not shown here). Thus, the non-collinearity could result from multi-spin higher order exchange. To test this hypothesis, we compare the exact field (black line) with the effective field related to the Heisenberg spin-Hamiltonian (red line) in Fig.~\ref{fig:Co:B}. Into the latter, the dynamically determined $J$'s enter.
There are clear deviations between the two fields, corroborating the presence of multi-spin exchange mechanisms. Unfortunately, there is no expression similar to the two-spin case \cite{Bruno2003,Streib2021,Solovyev2020} for multi-spin exchange interactions with constraining fields, which will be a focus of future studies.

\section{Summary and discussion\label{sec:Summary-and-discussion}}

In this work, we have considered adiabatic spin dynamics within a
tight-binding electronic structure theory based on constraining fields.
Furthermore, we have developed a method of extracting effective exchange
interactions $J_{ij}$ from the energy curvature tensor $\mathcal{J}_{ij}^{\alpha\beta}$
in non-collinear configurations. The effective exchange goes beyond
the simpler Heisenberg exchange, as it includes all contributions
to the two-spin exchange interaction up to infinite order, resulting
in an effective exchange interaction that depends on the magnetic
configuration. Within the spin dynamics simulations, we can track
the evolution of the effective exchange and its dependence on the
magnetic configuration.

In particular, we considered Ni, Fe, and Co dimers, and 
Fe and Co spin chains consisting of ten atoms each. The results show that
both moment lengths and effective exchange interactions depend dynamically on
the magnetic configuration. For strongly non-collinear states in particular, the results demonstrate a breakdown of a Heisenberg model description, which assumes
constant moment lengths and exchange interactions. In the case of
the Fe chain, the magnetic moments only change by a few percent
but the exchange interaction is strongly affected by non-collinearity,
with an increase of the average nearest-neighbor exchange by about
$10\;\%$ compared to the ferromagnetic ground state. For strongly non-collinear
states, the two-spin exchange interaction is insufficient to obtain the
correct effective field due to multi-spin interactions.

We have also shown that adiabatic spin dynamics at zero electronic temperature
without any additional damping conserves the energy but not the total
angular momentum. The adiabatic approximation implicitly introduces
a coupling of the electronic system to a bath in order to keep the
electrons in a quasi-equilibrium state. At zero temperature, heat
cannot be transferred to the bath, $\delta Q=T\delta S=0$, which
explains the conservation of energy despite the coupling to such a
bath. Angular momentum, however, can be transferred even at zero temperature
when the magnetic moment lengths are not constant. Our approach has
the advantage that it includes the change of moment lengths in non-collinear
configurations, which is a more accurate description of the physics
than assuming constant moments. In any real system, the spin dynamics
is coupled to the lattice, which allows a transfer of angular momentum
between the lattice and the magnetic moments. The disadvantage of
this adiabatic description is that it does not model this angular
momentum transfer on a microscopic level and only takes it into account
implicitly by a change of moment lengths without including the impact
on the lattice. Future work will be required to establish how angular
momentum conservation can be restored when including the lattice dynamics
within this adiabatic framework.

As a final remark, we note that the conservation laws analyzed here in Sec.~\ref{sec:Conservation-laws}, might have implications on how to interpret experimental  pump-probe experiments, such as the ones published in Ref.~\citep{Beaurepaire1996}. The dynamics of the angular momentum of the electron system and its transfer to a bath is, according to this analysis, distinctly different for collinear and non-collinear systems. Even if the experiment is made for a system that initially is collinear, say a ferromagnet, any transient excited state that has a non-collinear magnetic structure will open up for new channels of angular momentum transfer, that could be relevant for how to understand these types of experiments. Further studies of the model presented here will hopefully clarify this point. To this end, it might be necessary to simulate directly the electron dynamics without adiabatic approximation, see, e.g., recent works on ultra-fast spin dynamics based on tight-binding models \cite{Toepler2021,Hamamera2022}.

\input{Acknowledgement}

\appendix

\section{Derivative matrix\label{sec:Derivative-matrix}}

Naively, one would expect ${\partial e_{j\alpha}}/{\partial e_{j\beta}}=\delta_{\alpha\beta}$,
which is however not valid for a unit vector since the Cartesian components
are not independent. Instead, the derivative matrix 
evaluated at $\mathbf{e}_{j}=\sin\theta\,\mathbf{e}_{x}+\cos\theta\,\mathbf{e}_{z}$
is given by
\begin{equation}
\frac{\partial e_{j\alpha}}{\partial e_{j\beta}}=\begin{pmatrix}\cos^{2}\theta & 0 & -\sin\theta\cos\theta\\
0 & 1 & 0\\
-\sin\theta\cos\theta & 0 & \sin^{2}\theta
\end{pmatrix}_{\alpha\beta},
\end{equation}
which we obtain from the gradient of a unit vector in spherical coordinates,
\begin{equation}
\boldsymbol{\nabla}_{\mathbf{e}_{j}}e_{j\alpha}(\theta,\phi)=\frac{\partial e_{j\alpha}}{\partial\theta}\mathbf{e}_{\theta}+\frac{1}{\sin\theta}\frac{\partial e_{j\alpha}}{\partial\phi}\mathbf{e}_{\phi},
\end{equation}
with
\begin{equation}
\mathbf{e}_{j}=\begin{pmatrix}\sin\theta\cos\phi\\
\sin\theta\sin\phi\\
\cos\theta
\end{pmatrix},
\end{equation}
and
\begin{align}
\mathbf{e}_{\theta} & =\begin{pmatrix}\cos\theta\cos\phi\\
\cos\theta\sin\phi\\
-\sin\theta
\end{pmatrix},\;\mathbf{e}_{\phi}=\begin{pmatrix}-\sin\phi\\
\cos\phi\\
0
\end{pmatrix}.
\end{align}

\section{Dzyaloshinskii--Moriya interaction\label{sec:DMI}}

Here, we consider an additional DMI contribution to the generalized
exchange in Sec.~\ref{subsec:Generalized-exchange-interaction} \cite{Dzyaloshinsky1958,Moriya1960},
\begin{equation}
E_{\text{DMI}}=-\frac{1}{2}\sum_{ij}\mathbf{D}_{ij}\cdot\left(\mathbf{e}_{i}\times\mathbf{e}_{j}\right),
\end{equation}
with $\mathbf{D}_{ij}=-\mathbf{D}_{ji}$ and $\mathbf{D}_{ii}=0$. Such a contribution is allowed if spin-orbit coupling is taken into account and inversion symmetry is broken.
The DMI results in the following effective magnetic field in the reference
coordinate system (with $\mathbf{e}_{i}=\hat{\mathbf{z}}$ and $\mathbf{e}_{j}$
in the $xz$ plane),
\begin{align}
B_{ix}^{\text{eff}} & =\frac{1}{M_{i}}\sum_{j}\left(D_{ij}^{z}e_{jy}-D_{ij}^{y}e{}_{jz}\right),\\
B_{iy}^{\text{eff}} & =\frac{1}{M_{i}}\sum_{j}\left(-D_{ij}^{z}e_{jx}+D_{ij}^{x}e_{jz}\right),\\
B_{iz}^{\text{eff}} & =0.
\end{align}
Combining the effective exchange and this DMI, we obtain in the reference
coordinate system (for $i\neq j$)
\begin{eqnarray}
\mathcal{J}_{ij}^{xx} & = & J_{ij}(\theta_{ij})\cos^{2}\theta_{ij}\nonumber \\
 &  & +\sin\theta_{ij}\cos\theta_{ij}\left(J'_{ij}(\theta_{ij})+D_{ij}^{y}\right),\\
\mathcal{J}_{ij}^{xy} & = & D_{ij}^{z},\\
\mathcal{J}_{ij}^{xz} & = & -J_{ij}(\theta_{ij})\sin\theta_{ij}\cos\theta_{ij}\nonumber \\
 &  & -\sin^{2}\theta_{ij}\left(J'_{ij}(\theta_{ij})+D_{ij}^{y}\right),\\
\mathcal{J}_{ij}^{yx} & = & -D_{ij}^{z}\cos^{2}\theta_{ij}-D_{ij}^{x}\sin\theta_{ij}\cos\theta_{ij},\\
\mathcal{J}_{ij}^{yy} & = & J_{ij}(\theta_{ij}),\\
\mathcal{J}_{ij}^{yz} & = & D_{ij}^{x}\sin^{2}\theta_{ij}+D_{ij}^{z}\sin\theta_{ij}\cos\theta_{ij},\\
\mathcal{J}_{ij}^{z\beta} & = & 0.
\end{eqnarray}

We note that the contributions with $J'_{ij}(\theta_{ij})$ and $D_{ij}^{y}$
cannot be distinguished here from a single configuration and they both
change sign under exchange $i\leftrightarrow j$, see Eq.~(\ref{eq:J'}).
From a collinear state aligned along the $z$ axis, we can obtain the components $D_{ij}^{z}$ of the DMI,
in agreement with Ref.~\cite{Udvardi2003}.

\end{document}

%% file: Acknowledgement.tex
\section*{Acknowledgments}

This work was financially supported by the Knut and Alice Wallenberg Foundation through Grant No.\,2018.0060. O.E.~also acknowledges support by the Swedish Research Council (VR), the Foundation for Strategic Research (SSF), the Swedish Energy Agency (Energimyndigheten), the European Research Council (854843-FASTCORR), eSSENCE and STandUP. D.T.~and A.D.~acknowledge support from the Swedish Research Council (VR) with grant numbers VR 2016-05980, 2019-03666 and 2019-05304, respectively. The computations were enabled by resources provided by the Swedish National Infrastructure for Computing (SNIC) at the National Supercomputing Centre (NSC, Tetralith cluster), partially funded by the Swedish Research Council through Grant Agreements No. 2021-1-36 and No. 2021-5-395. We would like to thank Misha Katsnelson, Attila Szilva, and Pavel Bessarab for fruitful discussions.

%% file: main.bbl
\begin{thebibliography}{61}%
\makeatletter
\providecommand \@ifxundefined [1]{%
 \@ifx{#1\undefined}
}%
\providecommand \@ifnum [1]{%
 \ifnum #1\expandafter \@firstoftwo
 \else \expandafter \@secondoftwo
 \fi
}%
\providecommand \@ifx [1]{%
 \ifx #1\expandafter \@firstoftwo
 \else \expandafter \@secondoftwo
 \fi
}%
\providecommand \natexlab [1]{#1}%
\providecommand \enquote  [1]{``#1''}%
\providecommand \bibnamefont  [1]{#1}%
\providecommand \bibfnamefont [1]{#1}%
\providecommand \citenamefont [1]{#1}%
\providecommand \href@noop [0]{\@secondoftwo}%
\providecommand \href [0]{\begingroup \@sanitize@url \@href}%
\providecommand \@href[1]{\@@startlink{#1}\@@href}%
\providecommand \@@href[1]{\endgroup#1\@@endlink}%
\providecommand \@sanitize@url [0]{\catcode `\\12\catcode `\$12\catcode
  `\&12\catcode `\#12\catcode `\^12\catcode `\_12\catcode `\%12\relax}%
\providecommand \@@startlink[1]{}%
\providecommand \@@endlink[0]{}%
\providecommand \url  [0]{\begingroup\@sanitize@url \@url }%
\providecommand \@url [1]{\endgroup\@href {#1}{\urlprefix }}%
\providecommand \urlprefix  [0]{URL }%
\providecommand \Eprint [0]{\href }%
\providecommand \doibase [0]{https://doi.org/}%
\providecommand \selectlanguage [0]{\@gobble}%
\providecommand \bibinfo  [0]{\@secondoftwo}%
\providecommand \bibfield  [0]{\@secondoftwo}%
\providecommand \translation [1]{[#1]}%
\providecommand \BibitemOpen [0]{}%
\providecommand \bibitemStop [0]{}%
\providecommand \bibitemNoStop [0]{.\EOS\space}%
\providecommand \EOS [0]{\spacefactor3000\relax}%
\providecommand \BibitemShut  [1]{\csname bibitem#1\endcsname}%
\let\auto@bib@innerbib\@empty
\bibitem [{\citenamefont {Antropov}\ \emph {et~al.}(1995)\citenamefont
  {Antropov}, \citenamefont {Katsnelson}, \citenamefont {van Schilfgaarde},\
  and\ \citenamefont {Harmon}}]{Antropov1995}%
  \BibitemOpen
  \bibfield  {author} {\bibinfo {author} {\bibfnamefont {V.~P.}\ \bibnamefont
  {Antropov}}, \bibinfo {author} {\bibfnamefont {M.~I.}\ \bibnamefont
  {Katsnelson}}, \bibinfo {author} {\bibfnamefont {M.}~\bibnamefont {van
  Schilfgaarde}},\ and\ \bibinfo {author} {\bibfnamefont {B.~N.}\ \bibnamefont
  {Harmon}},\ }\bibfield  {title} {\bibinfo {title} {{$\mathit{Ab}$
  $\mathit{Initio}$ Spin Dynamics in Magnets}},\ }\href
  {https://doi.org/10.1103/PhysRevLett.75.729} {\bibfield  {journal} {\bibinfo
  {journal} {Phys. Rev. Lett.}\ }\textbf {\bibinfo {volume} {75}},\ \bibinfo
  {pages} {729} (\bibinfo {year} {1995})}\BibitemShut {NoStop}%
\bibitem [{\citenamefont {Antropov}\ \emph {et~al.}(1996)\citenamefont
  {Antropov}, \citenamefont {Katsnelson}, \citenamefont {Harmon}, \citenamefont
  {van Schilfgaarde},\ and\ \citenamefont {Kusnezov}}]{Antropov1996}%
  \BibitemOpen
  \bibfield  {author} {\bibinfo {author} {\bibfnamefont {V.~P.}\ \bibnamefont
  {Antropov}}, \bibinfo {author} {\bibfnamefont {M.~I.}\ \bibnamefont
  {Katsnelson}}, \bibinfo {author} {\bibfnamefont {B.~N.}\ \bibnamefont
  {Harmon}}, \bibinfo {author} {\bibfnamefont {M.}~\bibnamefont {van
  Schilfgaarde}},\ and\ \bibinfo {author} {\bibfnamefont {D.}~\bibnamefont
  {Kusnezov}},\ }\bibfield  {title} {\bibinfo {title} {Spin dynamics in
  magnets: Equation of motion and finite temperature effects},\ }\href
  {https://doi.org/10.1103/PhysRevB.54.1019} {\bibfield  {journal} {\bibinfo
  {journal} {Phys. Rev. B}\ }\textbf {\bibinfo {volume} {54}},\ \bibinfo
  {pages} {1019} (\bibinfo {year} {1996})}\BibitemShut {NoStop}%
\bibitem [{\citenamefont {Halilov}\ \emph {et~al.}(1998)\citenamefont
  {Halilov}, \citenamefont {Eschrig}, \citenamefont {Perlov},\ and\
  \citenamefont {Oppeneer}}]{Halilov1998}%
  \BibitemOpen
  \bibfield  {author} {\bibinfo {author} {\bibfnamefont {S.~V.}\ \bibnamefont
  {Halilov}}, \bibinfo {author} {\bibfnamefont {H.}~\bibnamefont {Eschrig}},
  \bibinfo {author} {\bibfnamefont {A.~Y.}\ \bibnamefont {Perlov}},\ and\
  \bibinfo {author} {\bibfnamefont {P.~M.}\ \bibnamefont {Oppeneer}},\
  }\bibfield  {title} {\bibinfo {title} {{Adiabatic spin dynamics from
  spin-density-functional theory: Application to Fe, Co, and Ni}},\ }\href
  {https://doi.org/10.1103/PhysRevB.58.293} {\bibfield  {journal} {\bibinfo
  {journal} {Phys. Rev. B}\ }\textbf {\bibinfo {volume} {58}},\ \bibinfo
  {pages} {293} (\bibinfo {year} {1998})}\BibitemShut {NoStop}%
\bibitem [{\citenamefont {Stocks}\ \emph {et~al.}(1998)\citenamefont {Stocks},
  \citenamefont {Ujfalussy}, \citenamefont {Wang}, \citenamefont {Nicholson},
  \citenamefont {Shelton}, \citenamefont {Wang}, \citenamefont {Canning},\ and\
  \citenamefont {Gy\"orffy}}]{Stocks1998}%
  \BibitemOpen
  \bibfield  {author} {\bibinfo {author} {\bibfnamefont {G.~M.}\ \bibnamefont
  {Stocks}}, \bibinfo {author} {\bibfnamefont {B.}~\bibnamefont {Ujfalussy}},
  \bibinfo {author} {\bibfnamefont {X.}~\bibnamefont {Wang}}, \bibinfo {author}
  {\bibfnamefont {D.~M.~C.}\ \bibnamefont {Nicholson}}, \bibinfo {author}
  {\bibfnamefont {W.~A.}\ \bibnamefont {Shelton}}, \bibinfo {author}
  {\bibfnamefont {Y.}~\bibnamefont {Wang}}, \bibinfo {author} {\bibfnamefont
  {A.}~\bibnamefont {Canning}},\ and\ \bibinfo {author} {\bibfnamefont {B.~L.}\
  \bibnamefont {Gy\"orffy}},\ }\bibfield  {title} {\bibinfo {title} {Towards a
  constrained local moment model for first principles spin dynamics},\ }\href
  {https://doi.org/10.1080/13642819808206775} {\bibfield  {journal} {\bibinfo
  {journal} {Philos. Mag. B}\ }\textbf {\bibinfo {volume} {78}},\ \bibinfo
  {pages} {665} (\bibinfo {year} {1998})}\BibitemShut {NoStop}%
\bibitem [{\citenamefont {Ujfalussy}\ \emph {et~al.}(1999)\citenamefont
  {Ujfalussy}, \citenamefont {Wang}, \citenamefont {Nicholson}, \citenamefont
  {Shelton}, \citenamefont {Stocks}, \citenamefont {Wang},\ and\ \citenamefont
  {Gyorffy}}]{Ujfalussy1999}%
  \BibitemOpen
  \bibfield  {author} {\bibinfo {author} {\bibfnamefont {B.}~\bibnamefont
  {Ujfalussy}}, \bibinfo {author} {\bibfnamefont {X.-D.}\ \bibnamefont {Wang}},
  \bibinfo {author} {\bibfnamefont {D.~M.~C.}\ \bibnamefont {Nicholson}},
  \bibinfo {author} {\bibfnamefont {W.~A.}\ \bibnamefont {Shelton}}, \bibinfo
  {author} {\bibfnamefont {G.~M.}\ \bibnamefont {Stocks}}, \bibinfo {author}
  {\bibfnamefont {Y.}~\bibnamefont {Wang}},\ and\ \bibinfo {author}
  {\bibfnamefont {B.~L.}\ \bibnamefont {Gyorffy}},\ }\bibfield  {title}
  {\bibinfo {title} {Constrained density functional theory for first principles
  spin dynamics},\ }\href {https://doi.org/10.1063/1.370494} {\bibfield
  {journal} {\bibinfo  {journal} {J. Appl. Phys.}\ }\textbf {\bibinfo {volume}
  {85}},\ \bibinfo {pages} {4824} (\bibinfo {year} {1999})}\BibitemShut
  {NoStop}%
\bibitem [{\citenamefont {Ma}\ and\ \citenamefont {Dudarev}(2015)}]{Ma2015}%
  \BibitemOpen
  \bibfield  {author} {\bibinfo {author} {\bibfnamefont {P.-W.}\ \bibnamefont
  {Ma}}\ and\ \bibinfo {author} {\bibfnamefont {S.~L.}\ \bibnamefont
  {Dudarev}},\ }\bibfield  {title} {\bibinfo {title} {Constrained density
  functional for noncollinear magnetism},\ }\href
  {https://doi.org/10.1103/PhysRevB.91.054420} {\bibfield  {journal} {\bibinfo
  {journal} {Phys. Rev. B}\ }\textbf {\bibinfo {volume} {91}},\ \bibinfo
  {pages} {054420} (\bibinfo {year} {2015})}\BibitemShut {NoStop}%
\bibitem [{\citenamefont {Singer}\ \emph {et~al.}(2005)\citenamefont {Singer},
  \citenamefont {F\"ahnle},\ and\ \citenamefont {Bihlmayer}}]{Singer2005}%
  \BibitemOpen
  \bibfield  {author} {\bibinfo {author} {\bibfnamefont {R.}~\bibnamefont
  {Singer}}, \bibinfo {author} {\bibfnamefont {M.}~\bibnamefont {F\"ahnle}},\
  and\ \bibinfo {author} {\bibfnamefont {G.}~\bibnamefont {Bihlmayer}},\
  }\bibfield  {title} {\bibinfo {title} {Constrained spin-density functional
  theory for excited magnetic configurations in an adiabatic approximation},\
  }\href {https://doi.org/10.1103/PhysRevB.71.214435} {\bibfield  {journal}
  {\bibinfo  {journal} {Phys. Rev. B}\ }\textbf {\bibinfo {volume} {71}},\
  \bibinfo {pages} {214435} (\bibinfo {year} {2005})}\BibitemShut {NoStop}%
\bibitem [{\citenamefont {Bertotti}\ \emph {et~al.}(2009)\citenamefont
  {Bertotti}, \citenamefont {Mayergoyz},\ and\ \citenamefont
  {Serpico}}]{Bertotti2009}%
  \BibitemOpen
  \bibfield  {author} {\bibinfo {author} {\bibfnamefont {G.}~\bibnamefont
  {Bertotti}}, \bibinfo {author} {\bibfnamefont {I.~D.}\ \bibnamefont
  {Mayergoyz}},\ and\ \bibinfo {author} {\bibfnamefont {C.}~\bibnamefont
  {Serpico}},\ }\href {https://doi.org/10.1016/B978-0-08-044316-4.X0001-1}
  {\emph {\bibinfo {title} {Nonlinear Magnetization Dynamics in Nanosystems}}}\
  (\bibinfo  {publisher} {Elsevier},\ \bibinfo {address} {Oxford},\ \bibinfo
  {year} {2009})\BibitemShut {NoStop}%
\bibitem [{\citenamefont {Eriksson}\ \emph {et~al.}(2017)\citenamefont
  {Eriksson}, \citenamefont {Bergman}, \citenamefont {Bergqvist},\ and\
  \citenamefont {Hellsvik}}]{Eriksson2017}%
  \BibitemOpen
  \bibfield  {author} {\bibinfo {author} {\bibfnamefont {O.}~\bibnamefont
  {Eriksson}}, \bibinfo {author} {\bibfnamefont {A.}~\bibnamefont {Bergman}},
  \bibinfo {author} {\bibfnamefont {L.}~\bibnamefont {Bergqvist}},\ and\
  \bibinfo {author} {\bibfnamefont {J.}~\bibnamefont {Hellsvik}},\ }\href
  {https://doi.org/10.1093/oso/9780198788669.001.0001} {\emph {\bibinfo {title}
  {Atomistic Spin Dynamics: Foundations and Applications}}}\ (\bibinfo
  {publisher} {Oxford University Press},\ \bibinfo {year} {2017})\BibitemShut
  {NoStop}%
\bibitem [{\citenamefont {Beaurepaire}\ \emph {et~al.}(1996)\citenamefont
  {Beaurepaire}, \citenamefont {Merle}, \citenamefont {Daunois},\ and\
  \citenamefont {Bigot}}]{Beaurepaire1996}%
  \BibitemOpen
  \bibfield  {author} {\bibinfo {author} {\bibfnamefont {E.}~\bibnamefont
  {Beaurepaire}}, \bibinfo {author} {\bibfnamefont {J.-C.}\ \bibnamefont
  {Merle}}, \bibinfo {author} {\bibfnamefont {A.}~\bibnamefont {Daunois}},\
  and\ \bibinfo {author} {\bibfnamefont {J.-Y.}\ \bibnamefont {Bigot}},\
  }\bibfield  {title} {\bibinfo {title} {Ultrafast spin dynamics in
  ferromagnetic nickel},\ }\href {https://doi.org/10.1103/PhysRevLett.76.4250}
  {\bibfield  {journal} {\bibinfo  {journal} {Phys. Rev. Lett.}\ }\textbf
  {\bibinfo {volume} {76}},\ \bibinfo {pages} {4250} (\bibinfo {year}
  {1996})}\BibitemShut {NoStop}%
\bibitem [{\citenamefont {B\"ottcher}\ \emph {et~al.}(2012)\citenamefont
  {B\"ottcher}, \citenamefont {Ernst},\ and\ \citenamefont
  {Henk}}]{Boettcher2012}%
  \BibitemOpen
  \bibfield  {author} {\bibinfo {author} {\bibfnamefont {D.}~\bibnamefont
  {B\"ottcher}}, \bibinfo {author} {\bibfnamefont {A.}~\bibnamefont {Ernst}},\
  and\ \bibinfo {author} {\bibfnamefont {J.}~\bibnamefont {Henk}},\ }\bibfield
  {title} {\bibinfo {title} {{Temperature-dependent Heisenberg exchange
  coupling constants from linking electronic-structure calculations and Monte
  Carlo simulations}},\ }\href
  {https://doi.org/https://doi.org/10.1016/j.jmmm.2011.08.053} {\bibfield
  {journal} {\bibinfo  {journal} {J. Magn. Magn. Mater.}\ }\textbf {\bibinfo
  {volume} {324}},\ \bibinfo {pages} {610} (\bibinfo {year}
  {2012})}\BibitemShut {NoStop}%
\bibitem [{\citenamefont {Mankovsky}\ \emph {et~al.}(2013)\citenamefont
  {Mankovsky}, \citenamefont {Polesya}, \citenamefont {Ebert}, \citenamefont
  {Bensch}, \citenamefont {Mathon}, \citenamefont {Pascarelli},\ and\
  \citenamefont {Min\'ar}}]{Mankovsky2013}%
  \BibitemOpen
  \bibfield  {author} {\bibinfo {author} {\bibfnamefont {S.}~\bibnamefont
  {Mankovsky}}, \bibinfo {author} {\bibfnamefont {S.}~\bibnamefont {Polesya}},
  \bibinfo {author} {\bibfnamefont {H.}~\bibnamefont {Ebert}}, \bibinfo
  {author} {\bibfnamefont {W.}~\bibnamefont {Bensch}}, \bibinfo {author}
  {\bibfnamefont {O.}~\bibnamefont {Mathon}}, \bibinfo {author} {\bibfnamefont
  {S.}~\bibnamefont {Pascarelli}},\ and\ \bibinfo {author} {\bibfnamefont
  {J.}~\bibnamefont {Min\'ar}},\ }\bibfield  {title} {\bibinfo {title}
  {{Pressure-induced bcc to hcp transition in Fe: Magnetism-driven structure
  transformation}},\ }\href {https://doi.org/10.1103/PhysRevB.88.184108}
  {\bibfield  {journal} {\bibinfo  {journal} {Phys. Rev. B}\ }\textbf {\bibinfo
  {volume} {88}},\ \bibinfo {pages} {184108} (\bibinfo {year}
  {2013})}\BibitemShut {NoStop}%
\bibitem [{\citenamefont {Mankovsky}\ \emph
  {et~al.}(2020{\natexlab{a}})\citenamefont {Mankovsky}, \citenamefont
  {Polesya},\ and\ \citenamefont {Ebert}}]{Mankovsky2020b}%
  \BibitemOpen
  \bibfield  {author} {\bibinfo {author} {\bibfnamefont {S.}~\bibnamefont
  {Mankovsky}}, \bibinfo {author} {\bibfnamefont {S.}~\bibnamefont {Polesya}},\
  and\ \bibinfo {author} {\bibfnamefont {H.}~\bibnamefont {Ebert}},\ }\bibfield
   {title} {\bibinfo {title} {Exchange coupling constants at finite
  temperature},\ }\href {https://doi.org/10.1103/PhysRevB.102.134434}
  {\bibfield  {journal} {\bibinfo  {journal} {Phys. Rev. B}\ }\textbf {\bibinfo
  {volume} {102}},\ \bibinfo {pages} {134434} (\bibinfo {year}
  {2020}{\natexlab{a}})}\BibitemShut {NoStop}%
\bibitem [{\citenamefont {R{\'{o}}zsa}\ \emph {et~al.}(2014)\citenamefont
  {R{\'{o}}zsa}, \citenamefont {Udvardi},\ and\ \citenamefont
  {Szunyogh}}]{Rozsa2014}%
  \BibitemOpen
  \bibfield  {author} {\bibinfo {author} {\bibfnamefont {L.}~\bibnamefont
  {R{\'{o}}zsa}}, \bibinfo {author} {\bibfnamefont {L.}~\bibnamefont
  {Udvardi}},\ and\ \bibinfo {author} {\bibfnamefont {L.}~\bibnamefont
  {Szunyogh}},\ }\bibfield  {title} {\bibinfo {title} {Langevin spin dynamics
  based on ab initio calculations: numerical schemes and applications},\ }\href
  {https://doi.org/10.1088/0953-8984/26/21/216003} {\bibfield  {journal}
  {\bibinfo  {journal} {J. Phys. Condens. Matter}\ }\textbf {\bibinfo {volume}
  {26}},\ \bibinfo {pages} {216003} (\bibinfo {year} {2014})}\BibitemShut
  {NoStop}%
\bibitem [{\citenamefont {Streib}\ \emph {et~al.}(2020)\citenamefont {Streib},
  \citenamefont {Borisov}, \citenamefont {Pereiro}, \citenamefont {Bergman},
  \citenamefont {Sj\"oqvist}, \citenamefont {Delin}, \citenamefont {Eriksson},\
  and\ \citenamefont {Thonig}}]{Streib2020}%
  \BibitemOpen
  \bibfield  {author} {\bibinfo {author} {\bibfnamefont {S.}~\bibnamefont
  {Streib}}, \bibinfo {author} {\bibfnamefont {V.}~\bibnamefont {Borisov}},
  \bibinfo {author} {\bibfnamefont {M.}~\bibnamefont {Pereiro}}, \bibinfo
  {author} {\bibfnamefont {A.}~\bibnamefont {Bergman}}, \bibinfo {author}
  {\bibfnamefont {E.}~\bibnamefont {Sj\"oqvist}}, \bibinfo {author}
  {\bibfnamefont {A.}~\bibnamefont {Delin}}, \bibinfo {author} {\bibfnamefont
  {O.}~\bibnamefont {Eriksson}},\ and\ \bibinfo {author} {\bibfnamefont
  {D.}~\bibnamefont {Thonig}},\ }\bibfield  {title} {\bibinfo {title} {Equation
  of motion and the constraining field in \textit{ab initio} spin dynamics},\
  }\href {https://doi.org/10.1103/PhysRevB.102.214407} {\bibfield  {journal}
  {\bibinfo  {journal} {Phys. Rev. B}\ }\textbf {\bibinfo {volume} {102}},\
  \bibinfo {pages} {214407} (\bibinfo {year} {2020})}\BibitemShut {NoStop}%
\bibitem [{\citenamefont {Cardias}\ \emph {et~al.}(2021)\citenamefont
  {Cardias}, \citenamefont {Barreteau}, \citenamefont {Thibaudeau},\ and\
  \citenamefont {Fu}}]{Cardias2021a}%
  \BibitemOpen
  \bibfield  {author} {\bibinfo {author} {\bibfnamefont {R.}~\bibnamefont
  {Cardias}}, \bibinfo {author} {\bibfnamefont {C.}~\bibnamefont {Barreteau}},
  \bibinfo {author} {\bibfnamefont {P.}~\bibnamefont {Thibaudeau}},\ and\
  \bibinfo {author} {\bibfnamefont {C.~C.}\ \bibnamefont {Fu}},\ }\bibfield
  {title} {\bibinfo {title} {Spin dynamics from a constrained magnetic
  tight-binding model},\ }\href {https://doi.org/10.1103/PhysRevB.103.235436}
  {\bibfield  {journal} {\bibinfo  {journal} {Phys. Rev. B}\ }\textbf {\bibinfo
  {volume} {103}},\ \bibinfo {pages} {235436} (\bibinfo {year}
  {2021})}\BibitemShut {NoStop}%
\bibitem [{\citenamefont {Streib}\ \emph {et~al.}(2021)\citenamefont {Streib},
  \citenamefont {Szilva}, \citenamefont {Borisov}, \citenamefont {Pereiro},
  \citenamefont {Bergman}, \citenamefont {Sj\"oqvist}, \citenamefont {Delin},
  \citenamefont {Katsnelson}, \citenamefont {Eriksson},\ and\ \citenamefont
  {Thonig}}]{Streib2021}%
  \BibitemOpen
  \bibfield  {author} {\bibinfo {author} {\bibfnamefont {S.}~\bibnamefont
  {Streib}}, \bibinfo {author} {\bibfnamefont {A.}~\bibnamefont {Szilva}},
  \bibinfo {author} {\bibfnamefont {V.}~\bibnamefont {Borisov}}, \bibinfo
  {author} {\bibfnamefont {M.}~\bibnamefont {Pereiro}}, \bibinfo {author}
  {\bibfnamefont {A.}~\bibnamefont {Bergman}}, \bibinfo {author} {\bibfnamefont
  {E.}~\bibnamefont {Sj\"oqvist}}, \bibinfo {author} {\bibfnamefont
  {A.}~\bibnamefont {Delin}}, \bibinfo {author} {\bibfnamefont {M.~I.}\
  \bibnamefont {Katsnelson}}, \bibinfo {author} {\bibfnamefont
  {O.}~\bibnamefont {Eriksson}},\ and\ \bibinfo {author} {\bibfnamefont
  {D.}~\bibnamefont {Thonig}},\ }\bibfield  {title} {\bibinfo {title}
  {{Exchange constants for local spin Hamiltonians from tight-binding
  models}},\ }\href {https://doi.org/10.1103/PhysRevB.103.224413} {\bibfield
  {journal} {\bibinfo  {journal} {Phys. Rev. B}\ }\textbf {\bibinfo {volume}
  {103}},\ \bibinfo {pages} {224413} (\bibinfo {year} {2021})}\BibitemShut
  {NoStop}%
\bibitem [{\citenamefont {Antropov}\ \emph {et~al.}(1997)\citenamefont
  {Antropov}, \citenamefont {Katsnelson},\ and\ \citenamefont
  {Liechtenstein}}]{Antropov1997}%
  \BibitemOpen
  \bibfield  {author} {\bibinfo {author} {\bibfnamefont {V.}~\bibnamefont
  {Antropov}}, \bibinfo {author} {\bibfnamefont {M.}~\bibnamefont
  {Katsnelson}},\ and\ \bibinfo {author} {\bibfnamefont {A.}~\bibnamefont
  {Liechtenstein}},\ }\bibfield  {title} {\bibinfo {title} {Exchange
  interactions in magnets},\ }\href
  {https://doi.org/10.1016/S0921-4526(97)00203-2} {\bibfield  {journal}
  {\bibinfo  {journal} {Physica B Condens. Matter}\ }\textbf {\bibinfo {volume}
  {237-238}},\ \bibinfo {pages} {336 } (\bibinfo {year} {1997})}\BibitemShut
  {NoStop}%
\bibitem [{\citenamefont {Antropov}\ \emph {et~al.}(1999)\citenamefont
  {Antropov}, \citenamefont {Harmon},\ and\ \citenamefont
  {Smirnov}}]{Antropov1999}%
  \BibitemOpen
  \bibfield  {author} {\bibinfo {author} {\bibfnamefont {V.}~\bibnamefont
  {Antropov}}, \bibinfo {author} {\bibfnamefont {B.}~\bibnamefont {Harmon}},\
  and\ \bibinfo {author} {\bibfnamefont {A.}~\bibnamefont {Smirnov}},\
  }\bibfield  {title} {\bibinfo {title} {Aspects of spin dynamics and magnetic
  interactions},\ }\href {https://doi.org/10.1016/S0304-8853(99)00425-4}
  {\bibfield  {journal} {\bibinfo  {journal} {J. Magn. Magn. Mater.}\ }\textbf
  {\bibinfo {volume} {200}},\ \bibinfo {pages} {148 } (\bibinfo {year}
  {1999})}\BibitemShut {NoStop}%
\bibitem [{\citenamefont {Szilva}\ \emph {et~al.}(2013)\citenamefont {Szilva},
  \citenamefont {Costa}, \citenamefont {Bergman}, \citenamefont {Szunyogh},
  \citenamefont {Nordstr\"om},\ and\ \citenamefont {Eriksson}}]{Szilva2013}%
  \BibitemOpen
  \bibfield  {author} {\bibinfo {author} {\bibfnamefont {A.}~\bibnamefont
  {Szilva}}, \bibinfo {author} {\bibfnamefont {M.}~\bibnamefont {Costa}},
  \bibinfo {author} {\bibfnamefont {A.}~\bibnamefont {Bergman}}, \bibinfo
  {author} {\bibfnamefont {L.}~\bibnamefont {Szunyogh}}, \bibinfo {author}
  {\bibfnamefont {L.}~\bibnamefont {Nordstr\"om}},\ and\ \bibinfo {author}
  {\bibfnamefont {O.}~\bibnamefont {Eriksson}},\ }\bibfield  {title} {\bibinfo
  {title} {{Interatomic Exchange Interactions for Finite-Temperature Magnetism
  and Nonequilibrium Spin Dynamics}},\ }\href
  {https://doi.org/10.1103/PhysRevLett.111.127204} {\bibfield  {journal}
  {\bibinfo  {journal} {Phys. Rev. Lett.}\ }\textbf {\bibinfo {volume} {111}},\
  \bibinfo {pages} {127204} (\bibinfo {year} {2013})}\BibitemShut {NoStop}%
\bibitem [{\citenamefont {Secchi}\ \emph {et~al.}(2015)\citenamefont {Secchi},
  \citenamefont {Lichtenstein},\ and\ \citenamefont {Katsnelson}}]{Secchi2015}%
  \BibitemOpen
  \bibfield  {author} {\bibinfo {author} {\bibfnamefont {A.}~\bibnamefont
  {Secchi}}, \bibinfo {author} {\bibfnamefont {A.}~\bibnamefont
  {Lichtenstein}},\ and\ \bibinfo {author} {\bibfnamefont {M.}~\bibnamefont
  {Katsnelson}},\ }\bibfield  {title} {\bibinfo {title} {Magnetic interactions
  in strongly correlated systems: Spin and orbital contributions},\ }\href
  {https://doi.org/https://doi.org/10.1016/j.aop.2015.05.002} {\bibfield
  {journal} {\bibinfo  {journal} {Ann. Phys. (N. Y.)}\ }\textbf {\bibinfo
  {volume} {360}},\ \bibinfo {pages} {61 } (\bibinfo {year}
  {2015})}\BibitemShut {NoStop}%
\bibitem [{\citenamefont {Szilva}\ \emph {et~al.}(2017)\citenamefont {Szilva},
  \citenamefont {Thonig}, \citenamefont {Bessarab}, \citenamefont {Kvashnin},
  \citenamefont {Rodrigues}, \citenamefont {Cardias}, \citenamefont {Pereiro},
  \citenamefont {Nordstr\"om}, \citenamefont {Bergman}, \citenamefont
  {Klautau},\ and\ \citenamefont {Eriksson}}]{Szilva2017}%
  \BibitemOpen
  \bibfield  {author} {\bibinfo {author} {\bibfnamefont {A.}~\bibnamefont
  {Szilva}}, \bibinfo {author} {\bibfnamefont {D.}~\bibnamefont {Thonig}},
  \bibinfo {author} {\bibfnamefont {P.~F.}\ \bibnamefont {Bessarab}}, \bibinfo
  {author} {\bibfnamefont {Y.~O.}\ \bibnamefont {Kvashnin}}, \bibinfo {author}
  {\bibfnamefont {D.~C.~M.}\ \bibnamefont {Rodrigues}}, \bibinfo {author}
  {\bibfnamefont {R.}~\bibnamefont {Cardias}}, \bibinfo {author} {\bibfnamefont
  {M.}~\bibnamefont {Pereiro}}, \bibinfo {author} {\bibfnamefont
  {L.}~\bibnamefont {Nordstr\"om}}, \bibinfo {author} {\bibfnamefont
  {A.}~\bibnamefont {Bergman}}, \bibinfo {author} {\bibfnamefont {A.~B.}\
  \bibnamefont {Klautau}},\ and\ \bibinfo {author} {\bibfnamefont
  {O.}~\bibnamefont {Eriksson}},\ }\bibfield  {title} {\bibinfo {title}
  {{Theory of noncollinear interactions beyond Heisenberg exchange:
  Applications to bcc Fe}},\ }\href
  {https://doi.org/10.1103/PhysRevB.96.144413} {\bibfield  {journal} {\bibinfo
  {journal} {Phys. Rev. B}\ }\textbf {\bibinfo {volume} {96}},\ \bibinfo
  {pages} {144413} (\bibinfo {year} {2017})}\BibitemShut {NoStop}%
\bibitem [{\citenamefont {Cardias}\ \emph
  {et~al.}(2020{\natexlab{a}})\citenamefont {Cardias}, \citenamefont {Szilva},
  \citenamefont {Bezerra-Neto}, \citenamefont {Ribeiro}, \citenamefont
  {Bergman}, \citenamefont {Kvashnin}, \citenamefont {Fransson}, \citenamefont
  {Klautau}, \citenamefont {Eriksson},\ and\ \citenamefont
  {Nordstr{\"o}m}}]{Cardias2020a}%
  \BibitemOpen
  \bibfield  {author} {\bibinfo {author} {\bibfnamefont {R.}~\bibnamefont
  {Cardias}}, \bibinfo {author} {\bibfnamefont {A.}~\bibnamefont {Szilva}},
  \bibinfo {author} {\bibfnamefont {M.~M.}\ \bibnamefont {Bezerra-Neto}},
  \bibinfo {author} {\bibfnamefont {M.~S.}\ \bibnamefont {Ribeiro}}, \bibinfo
  {author} {\bibfnamefont {A.}~\bibnamefont {Bergman}}, \bibinfo {author}
  {\bibfnamefont {Y.~O.}\ \bibnamefont {Kvashnin}}, \bibinfo {author}
  {\bibfnamefont {J.}~\bibnamefont {Fransson}}, \bibinfo {author}
  {\bibfnamefont {A.~B.}\ \bibnamefont {Klautau}}, \bibinfo {author}
  {\bibfnamefont {O.}~\bibnamefont {Eriksson}},\ and\ \bibinfo {author}
  {\bibfnamefont {L.}~\bibnamefont {Nordstr{\"o}m}},\ }\bibfield  {title}
  {\bibinfo {title} {{First-principles Dzyaloshinskii--Moriya interaction in a
  non-collinear framework}},\ }\href
  {https://doi.org/10.1038/s41598-020-77219-3} {\bibfield  {journal} {\bibinfo
  {journal} {Sci. Rep.}\ }\textbf {\bibinfo {volume} {10}},\ \bibinfo {pages}
  {20339} (\bibinfo {year} {2020}{\natexlab{a}})}\BibitemShut {NoStop}%
\bibitem [{\citenamefont {Cardias}\ \emph
  {et~al.}(2020{\natexlab{b}})\citenamefont {Cardias}, \citenamefont {Bergman},
  \citenamefont {Szilva}, \citenamefont {Kvashnin}, \citenamefont {Fransson},
  \citenamefont {Klautau}, \citenamefont {Eriksson},\ and\ \citenamefont
  {Nordstr\"om}}]{Cardias2020b}%
  \BibitemOpen
  \bibfield  {author} {\bibinfo {author} {\bibfnamefont {R.}~\bibnamefont
  {Cardias}}, \bibinfo {author} {\bibfnamefont {A.}~\bibnamefont {Bergman}},
  \bibinfo {author} {\bibfnamefont {A.}~\bibnamefont {Szilva}}, \bibinfo
  {author} {\bibfnamefont {Y.~O.}\ \bibnamefont {Kvashnin}}, \bibinfo {author}
  {\bibfnamefont {J.}~\bibnamefont {Fransson}}, \bibinfo {author}
  {\bibfnamefont {A.~B.}\ \bibnamefont {Klautau}}, \bibinfo {author}
  {\bibfnamefont {O.}~\bibnamefont {Eriksson}},\ and\ \bibinfo {author}
  {\bibfnamefont {L.}~\bibnamefont {Nordstr\"om}},\ }\href@noop {} {\bibinfo
  {title} {{Dzyaloshinskii-Moriya interaction in absence of spin-orbit
  coupling}}} (\bibinfo {year} {2020}{\natexlab{b}}),\ \Eprint
  {https://arxiv.org/abs/arXiv:2003.04680} {arXiv:2003.04680} \BibitemShut
  {NoStop}%
\bibitem [{\citenamefont {Tanaka}\ and\ \citenamefont
  {Ury\^u}(1977)}]{Tanaka1977}%
  \BibitemOpen
  \bibfield  {author} {\bibinfo {author} {\bibfnamefont {Y.}~\bibnamefont
  {Tanaka}}\ and\ \bibinfo {author} {\bibfnamefont {N.}~\bibnamefont {Ury\^u}},\
  }\bibfield  {title} {\bibinfo {title} {{Exchange Interactions in
  Antiferromagnetic FeI2. II. The Four Spin Interaction}},\ }\href
  {https://doi.org/10.1143/JPSJ.43.1569} {\bibfield  {journal} {\bibinfo
  {journal} {J. Phys. Soc. Jpn.}\ }\textbf {\bibinfo {volume} {43}},\ \bibinfo
  {pages} {1569} (\bibinfo {year} {1977})}\BibitemShut {NoStop}%
\bibitem [{\citenamefont {MacDonald}\ \emph {et~al.}(1988)\citenamefont
  {MacDonald}, \citenamefont {Girvin},\ and\ \citenamefont
  {Yoshioka}}]{MacDonald1988}%
  \BibitemOpen
  \bibfield  {author} {\bibinfo {author} {\bibfnamefont {A.~H.}\ \bibnamefont
  {MacDonald}}, \bibinfo {author} {\bibfnamefont {S.~M.}\ \bibnamefont
  {Girvin}},\ and\ \bibinfo {author} {\bibfnamefont {D.}~\bibnamefont
  {Yoshioka}},\ }\bibfield  {title} {\bibinfo {title} {{$\frac{t}{U}$ expansion
  for the Hubbard model}},\ }\href {https://doi.org/10.1103/PhysRevB.37.9753}
  {\bibfield  {journal} {\bibinfo  {journal} {Phys. Rev. B}\ }\textbf {\bibinfo
  {volume} {37}},\ \bibinfo {pages} {9753} (\bibinfo {year}
  {1988})}\BibitemShut {NoStop}%
\bibitem [{\citenamefont {Singer}\ \emph {et~al.}(2011)\citenamefont {Singer},
  \citenamefont {Dietermann},\ and\ \citenamefont {F\"ahnle}}]{Singer2011}%
  \BibitemOpen
  \bibfield  {author} {\bibinfo {author} {\bibfnamefont {R.}~\bibnamefont
  {Singer}}, \bibinfo {author} {\bibfnamefont {F.}~\bibnamefont {Dietermann}},\
  and\ \bibinfo {author} {\bibfnamefont {M.}~\bibnamefont {F\"ahnle}},\
  }\bibfield  {title} {\bibinfo {title} {{Spin Interactions in bcc and fcc Fe
  beyond the Heisenberg Model}},\ }\href
  {https://doi.org/10.1103/PhysRevLett.107.017204} {\bibfield  {journal}
  {\bibinfo  {journal} {Phys. Rev. Lett.}\ }\textbf {\bibinfo {volume} {107}},\
  \bibinfo {pages} {017204} (\bibinfo {year} {2011})}\BibitemShut {NoStop}%
\bibitem [{\citenamefont {Hoffmann}\ and\ \citenamefont
  {Bl\"ugel}(2020)}]{Hoffmann2020}%
  \BibitemOpen
  \bibfield  {author} {\bibinfo {author} {\bibfnamefont {M.}~\bibnamefont
  {Hoffmann}}\ and\ \bibinfo {author} {\bibfnamefont {S.}~\bibnamefont
  {Bl\"ugel}},\ }\bibfield  {title} {\bibinfo {title} {{Systematic derivation
  of realistic spin models for beyond-Heisenberg solids}},\ }\href
  {https://doi.org/10.1103/PhysRevB.101.024418} {\bibfield  {journal} {\bibinfo
   {journal} {Phys. Rev. B}\ }\textbf {\bibinfo {volume} {101}},\ \bibinfo
  {pages} {024418} (\bibinfo {year} {2020})}\BibitemShut {NoStop}%
\bibitem [{\citenamefont {Brinker}\ \emph {et~al.}(2020)\citenamefont
  {Brinker}, \citenamefont {dos Santos~Dias},\ and\ \citenamefont
  {Lounis}}]{Brinker2020}%
  \BibitemOpen
  \bibfield  {author} {\bibinfo {author} {\bibfnamefont {S.}~\bibnamefont
  {Brinker}}, \bibinfo {author} {\bibfnamefont {M.}~\bibnamefont {dos
  Santos~Dias}},\ and\ \bibinfo {author} {\bibfnamefont {S.}~\bibnamefont
  {Lounis}},\ }\bibfield  {title} {\bibinfo {title} {Prospecting chiral
  multisite interactions in prototypical magnetic systems},\ }\href
  {https://doi.org/10.1103/PhysRevResearch.2.033240} {\bibfield  {journal}
  {\bibinfo  {journal} {Phys. Rev. Research}\ }\textbf {\bibinfo {volume}
  {2}},\ \bibinfo {pages} {033240} (\bibinfo {year} {2020})}\BibitemShut
  {NoStop}%
\bibitem [{\citenamefont {Mankovsky}\ \emph
  {et~al.}(2020{\natexlab{b}})\citenamefont {Mankovsky}, \citenamefont
  {Polesya},\ and\ \citenamefont {Ebert}}]{Mankovsky2020}%
  \BibitemOpen
  \bibfield  {author} {\bibinfo {author} {\bibfnamefont {S.}~\bibnamefont
  {Mankovsky}}, \bibinfo {author} {\bibfnamefont {S.}~\bibnamefont {Polesya}},\
  and\ \bibinfo {author} {\bibfnamefont {H.}~\bibnamefont {Ebert}},\ }\bibfield
   {title} {\bibinfo {title} {{Extension of the standard Heisenberg Hamiltonian
  to multispin exchange interactions}},\ }\href
  {https://doi.org/10.1103/PhysRevB.101.174401} {\bibfield  {journal} {\bibinfo
   {journal} {Phys. Rev. B}\ }\textbf {\bibinfo {volume} {101}},\ \bibinfo
  {pages} {174401} (\bibinfo {year} {2020}{\natexlab{b}})}\BibitemShut
  {NoStop}%
\bibitem [{\citenamefont {Antropov}(2003)}]{Antropov2003}%
  \BibitemOpen
  \bibfield  {author} {\bibinfo {author} {\bibfnamefont {V.}~\bibnamefont
  {Antropov}},\ }\bibfield  {title} {\bibinfo {title} {The exchange coupling
  and spin waves in metallic magnets: removal of the long-wave approximation},\
  }\href {https://doi.org/https://doi.org/10.1016/S0304-8853(03)00206-3}
  {\bibfield  {journal} {\bibinfo  {journal} {J. Magn. Magn. Mater.}\ }\textbf
  {\bibinfo {volume} {262}},\ \bibinfo {pages} {L192 } (\bibinfo {year}
  {2003})}\BibitemShut {NoStop}%
\bibitem [{\citenamefont {Katsnelson}\ and\ \citenamefont
  {Lichtenstein}(2004)}]{Katsnelson2004}%
  \BibitemOpen
  \bibfield  {author} {\bibinfo {author} {\bibfnamefont {M.~I.}\ \bibnamefont
  {Katsnelson}}\ and\ \bibinfo {author} {\bibfnamefont {A.~I.}\ \bibnamefont
  {Lichtenstein}},\ }\bibfield  {title} {\bibinfo {title} {Magnetic
  susceptibility, exchange interactions and spin-wave spectra in the local spin
  density approximation},\ }\href {https://doi.org/10.1088/0953-8984/16/41/023}
  {\bibfield  {journal} {\bibinfo  {journal} {J. Phys. Condens. Matter}\
  }\textbf {\bibinfo {volume} {16}},\ \bibinfo {pages} {7439} (\bibinfo {year}
  {2004})}\BibitemShut {NoStop}%
\bibitem [{\citenamefont {Buczek}\ \emph {et~al.}(2011)\citenamefont {Buczek},
  \citenamefont {Ernst},\ and\ \citenamefont {Sandratskii}}]{Buczek2011}%
  \BibitemOpen
  \bibfield  {author} {\bibinfo {author} {\bibfnamefont {P.}~\bibnamefont
  {Buczek}}, \bibinfo {author} {\bibfnamefont {A.}~\bibnamefont {Ernst}},\ and\
  \bibinfo {author} {\bibfnamefont {L.~M.}\ \bibnamefont {Sandratskii}},\
  }\bibfield  {title} {\bibinfo {title} {Different dimensionality trends in the
  Landau damping of magnons in iron, cobalt, and nickel: Time-dependent density
  functional study},\ }\href {https://doi.org/10.1103/PhysRevB.84.174418}
  {\bibfield  {journal} {\bibinfo  {journal} {Phys. Rev. B}\ }\textbf {\bibinfo
  {volume} {84}},\ \bibinfo {pages} {174418} (\bibinfo {year}
  {2011})}\BibitemShut {NoStop}%
\bibitem [{\citenamefont {Durhuus}\ \emph {et~al.}(2022)\citenamefont
  {Durhuus}, \citenamefont {Skovhus},\ and\ \citenamefont
  {Olsen}}]{Durhuus2022}%
  \BibitemOpen
  \bibfield  {author} {\bibinfo {author} {\bibfnamefont {F.}~\bibnamefont
  {Durhuus}}, \bibinfo {author} {\bibfnamefont {T.}~\bibnamefont {Skovhus}},\
  and\ \bibinfo {author} {\bibfnamefont {T.}~\bibnamefont {Olsen}},\
  }\href@noop {} {\bibinfo {title} {Plane wave implementation of the magnetic
  force theorem for magnetic exchange constants: Application to bulk Fe, Co and
  Ni}} (\bibinfo {year} {2022}),\ \Eprint
  {https://arxiv.org/abs/arXiv:2204.04169} {arXiv:2204.04169} \BibitemShut
  {NoStop}%
\bibitem [{\citenamefont {Landau}\ and\ \citenamefont
  {Lifshitz}(1977)}]{LandauLifshitz3}%
  \BibitemOpen
  \bibfield  {author} {\bibinfo {author} {\bibfnamefont {L.~D.}\ \bibnamefont
  {Landau}}\ and\ \bibinfo {author} {\bibfnamefont {E.~M.}\ \bibnamefont
  {Lifshitz}},\ }\href@noop {} {\emph {\bibinfo {title} {Quantum Mechanics,
  Non-Relativistic Theory}}},\ \bibinfo {edition} {3rd}\ ed.\ (\bibinfo
  {publisher} {Pergamon Press, Oxford},\ \bibinfo {year} {1977})\BibitemShut
  {NoStop}%
\bibitem [{\citenamefont {T\"ows}\ and\ \citenamefont
  {Pastor}(2015)}]{Toews2015}%
  \BibitemOpen
  \bibfield  {author} {\bibinfo {author} {\bibfnamefont {W.}~\bibnamefont
  {T\"ows}}\ and\ \bibinfo {author} {\bibfnamefont {G.~M.}\ \bibnamefont
  {Pastor}},\ }\bibfield  {title} {\bibinfo {title} {Many-body theory of
  ultrafast demagnetization and angular momentum transfer in ferromagnetic
  transition metals},\ }\href {https://doi.org/10.1103/PhysRevLett.115.217204}
  {\bibfield  {journal} {\bibinfo  {journal} {Phys. Rev. Lett.}\ }\textbf
  {\bibinfo {volume} {115}},\ \bibinfo {pages} {217204} (\bibinfo {year}
  {2015})}\BibitemShut {NoStop}%
\bibitem [{\citenamefont {Gilbert}(2004)}]{Gilbert2004}%
  \BibitemOpen
  \bibfield  {author} {\bibinfo {author} {\bibfnamefont {T.~L.}\ \bibnamefont
  {Gilbert}},\ }\bibfield  {title} {\bibinfo {title} {A phenomenological theory
  of damping in ferromagnetic materials},\ }\href
  {https://doi.org/10.1109/TMAG.2004.836740} {\bibfield  {journal} {\bibinfo
  {journal} {IEEE Trans. Magn.}\ }\textbf {\bibinfo {volume} {40}},\ \bibinfo
  {pages} {3443} (\bibinfo {year} {2004})}\BibitemShut {NoStop}%
\bibitem [{\citenamefont {Mentink}\ \emph {et~al.}(2010)\citenamefont
  {Mentink}, \citenamefont {Tretyakov}, \citenamefont {Fasolino}, \citenamefont
  {Katsnelson},\ and\ \citenamefont {Rasing}}]{Mentink2010}%
  \BibitemOpen
  \bibfield  {author} {\bibinfo {author} {\bibfnamefont {J.~H.}\ \bibnamefont
  {Mentink}}, \bibinfo {author} {\bibfnamefont {M.~V.}\ \bibnamefont
  {Tretyakov}}, \bibinfo {author} {\bibfnamefont {A.}~\bibnamefont {Fasolino}},
  \bibinfo {author} {\bibfnamefont {M.~I.}\ \bibnamefont {Katsnelson}},\ and\
  \bibinfo {author} {\bibfnamefont {T.}~\bibnamefont {Rasing}},\ }\bibfield
  {title} {\bibinfo {title} {{Stable and fast semi-implicit integration of the
  stochastic Landau{\textendash}Lifshitz equation}},\ }\href
  {https://doi.org/10.1088/0953-8984/22/17/176001} {\bibfield  {journal}
  {\bibinfo  {journal} {J. Phys. Condens. Matter}\ }\textbf {\bibinfo {volume}
  {22}},\ \bibinfo {pages} {176001} (\bibinfo {year} {2010})}\BibitemShut
  {NoStop}%
\bibitem [{\citenamefont {Yuan}\ \emph {et~al.}(2014)\citenamefont {Yuan},
  \citenamefont {Hals}, \citenamefont {Liu}, \citenamefont {Starikov},
  \citenamefont {Brataas},\ and\ \citenamefont {Kelly}}]{Yuan2014}%
  \BibitemOpen
  \bibfield  {author} {\bibinfo {author} {\bibfnamefont {Z.}~\bibnamefont
  {Yuan}}, \bibinfo {author} {\bibfnamefont {K.~M.~D.}\ \bibnamefont {Hals}},
  \bibinfo {author} {\bibfnamefont {Y.}~\bibnamefont {Liu}}, \bibinfo {author}
  {\bibfnamefont {A.~A.}\ \bibnamefont {Starikov}}, \bibinfo {author}
  {\bibfnamefont {A.}~\bibnamefont {Brataas}},\ and\ \bibinfo {author}
  {\bibfnamefont {P.~J.}\ \bibnamefont {Kelly}},\ }\bibfield  {title} {\bibinfo
  {title} {Gilbert damping in noncollinear ferromagnets},\ }\href
  {https://doi.org/10.1103/PhysRevLett.113.266603} {\bibfield  {journal}
  {\bibinfo  {journal} {Phys. Rev. Lett.}\ }\textbf {\bibinfo {volume} {113}},\
  \bibinfo {pages} {266603} (\bibinfo {year} {2014})}\BibitemShut {NoStop}%
\bibitem [{\citenamefont {Mankovsky}\ \emph {et~al.}(2018)\citenamefont
  {Mankovsky}, \citenamefont {Wimmer},\ and\ \citenamefont
  {Ebert}}]{Mankovsky2018}%
  \BibitemOpen
  \bibfield  {author} {\bibinfo {author} {\bibfnamefont {S.}~\bibnamefont
  {Mankovsky}}, \bibinfo {author} {\bibfnamefont {S.}~\bibnamefont {Wimmer}},\
  and\ \bibinfo {author} {\bibfnamefont {H.}~\bibnamefont {Ebert}},\ }\bibfield
   {title} {\bibinfo {title} {Gilbert damping in noncollinear magnetic
  systems},\ }\href {https://doi.org/10.1103/PhysRevB.98.104406} {\bibfield
  {journal} {\bibinfo  {journal} {Phys. Rev. B}\ }\textbf {\bibinfo {volume}
  {98}},\ \bibinfo {pages} {104406} (\bibinfo {year} {2018})}\BibitemShut
  {NoStop}%
\bibitem [{\citenamefont {Brinker}\ \emph {et~al.}(2022)\citenamefont
  {Brinker}, \citenamefont {dos Santos~Dias},\ and\ \citenamefont
  {Lounis}}]{Brinker2022}%
  \BibitemOpen
  \bibfield  {author} {\bibinfo {author} {\bibfnamefont {S.}~\bibnamefont
  {Brinker}}, \bibinfo {author} {\bibfnamefont {M.}~\bibnamefont {dos
  Santos~Dias}},\ and\ \bibinfo {author} {\bibfnamefont {S.}~\bibnamefont
  {Lounis}},\ }\href@noop {} {\bibinfo {title} {{Generalization of the
  Landau-Lifshitz-Gilbert equation by multi-body contributions to Gilbert
  damping for non-collinear magnets}}} (\bibinfo {year} {2022}),\ \Eprint
  {https://arxiv.org/abs/arXiv:2202.06154} {arXiv:2202.06154} \BibitemShut
  {NoStop}%
\bibitem [{\citenamefont {Udvardi}\ \emph {et~al.}(2003)\citenamefont
  {Udvardi}, \citenamefont {Szunyogh}, \citenamefont {Palot\'as},\ and\
  \citenamefont {Weinberger}}]{Udvardi2003}%
  \BibitemOpen
  \bibfield  {author} {\bibinfo {author} {\bibfnamefont {L.}~\bibnamefont
  {Udvardi}}, \bibinfo {author} {\bibfnamefont {L.}~\bibnamefont {Szunyogh}},
  \bibinfo {author} {\bibfnamefont {K.}~\bibnamefont {Palot\'as}},\ and\
  \bibinfo {author} {\bibfnamefont {P.}~\bibnamefont {Weinberger}},\ }\bibfield
   {title} {\bibinfo {title} {First-principles relativistic study of spin waves
  in thin magnetic films},\ }\href {https://doi.org/10.1103/PhysRevB.68.104436}
  {\bibfield  {journal} {\bibinfo  {journal} {Phys. Rev. B}\ }\textbf {\bibinfo
  {volume} {68}},\ \bibinfo {pages} {104436} (\bibinfo {year}
  {2003})}\BibitemShut {NoStop}%
\bibitem [{\citenamefont {Dzyaloshinsky}(1958)}]{Dzyaloshinsky1958}%
  \BibitemOpen
  \bibfield  {author} {\bibinfo {author} {\bibfnamefont {I.}~\bibnamefont
  {Dzyaloshinsky}},\ }\bibfield  {title} {\bibinfo {title} {A thermodynamic
  theory of ``weak'' ferromagnetism of antiferromagnetics},\ }\href
  {https://doi.org/https://doi.org/10.1016/0022-3697(58)90076-3} {\bibfield
  {journal} {\bibinfo  {journal} {J. Phys. Chem. Solids}\ }\textbf {\bibinfo
  {volume} {4}},\ \bibinfo {pages} {241} (\bibinfo {year} {1958})}\BibitemShut
  {NoStop}%
\bibitem [{\citenamefont {Moriya}(1960)}]{Moriya1960}%
  \BibitemOpen
  \bibfield  {author} {\bibinfo {author} {\bibfnamefont {T.}~\bibnamefont
  {Moriya}},\ }\bibfield  {title} {\bibinfo {title} {Anisotropic superexchange
  interaction and weak ferromagnetism},\ }\href
  {https://doi.org/10.1103/PhysRev.120.91} {\bibfield  {journal} {\bibinfo
  {journal} {Phys. Rev.}\ }\textbf {\bibinfo {volume} {120}},\ \bibinfo {pages}
  {91} (\bibinfo {year} {1960})}\BibitemShut {NoStop}%
\bibitem [{CAH()}]{CAHMD}%
  \BibitemOpen
  \href@noop {} {}\bibinfo {note} {{Computer code CAHMD, classical atomistic
  hybrid multi-degree dynamics. A computer program package for atomistic
  dynamics simulations of multiple degrees of freedom (e.g. electron,
  magnetization, lattice vibrations) based on parametrized Hamiltonians. (Danny
  Thonig, danny.thonig@oru.se, 2013) (unpublished, available from
  \url{https://cahmd.gitlab.io/cahmdweb/})}}\BibitemShut {NoStop}%
\bibitem [{\citenamefont {Aut{\`{e}}s}\ \emph {et~al.}(2006)\citenamefont
  {Aut{\`{e}}s}, \citenamefont {Barreteau}, \citenamefont {Spanjaard},\ and\
  \citenamefont {Desjonqu{\`{e}}res}}]{Aute2006}%
  \BibitemOpen
  \bibfield  {author} {\bibinfo {author} {\bibfnamefont {G.}~\bibnamefont
  {Aut{\`{e}}s}}, \bibinfo {author} {\bibfnamefont {C.}~\bibnamefont
  {Barreteau}}, \bibinfo {author} {\bibfnamefont {D.}~\bibnamefont
  {Spanjaard}},\ and\ \bibinfo {author} {\bibfnamefont {M.-C.}\ \bibnamefont
  {Desjonqu{\`{e}}res}},\ }\bibfield  {title} {\bibinfo {title} {Magnetism of
  iron: from the bulk to the monatomic wire},\ }\href
  {https://doi.org/10.1088/0953-8984/18/29/018} {\bibfield  {journal} {\bibinfo
   {journal} {J. Phys. Condens. Matter}\ }\textbf {\bibinfo {volume} {18}},\
  \bibinfo {pages} {6785} (\bibinfo {year} {2006})}\BibitemShut {NoStop}%
\bibitem [{\citenamefont {Rossen}(2019)}]{Rossen2019}%
  \BibitemOpen
  \bibfield  {author} {\bibinfo {author} {\bibfnamefont {S.}~\bibnamefont
  {Rossen}},\ }\emph {\bibinfo {title} {{Magnetization disorder at finite
  temperature: A tight-binding Monte Carlo modelling and spin dynamics study of
  bulk iron and cobalt clusters: theory, numerical implementation and
  simulations}}},\ \href {https://hdl.handle.net/2066/203851} {Ph.D. thesis},\
  \bibinfo  {school} {Radboud University Nijmegen} (\bibinfo {year}
  {2019})\BibitemShut {NoStop}%
\bibitem [{\citenamefont {Slater}\ and\ \citenamefont
  {Koster}(1954)}]{Slater1954}%
  \BibitemOpen
  \bibfield  {author} {\bibinfo {author} {\bibfnamefont {J.~C.}\ \bibnamefont
  {Slater}}\ and\ \bibinfo {author} {\bibfnamefont {G.~F.}\ \bibnamefont
  {Koster}},\ }\bibfield  {title} {\bibinfo {title} {{Simplified LCAO Method
  for the Periodic Potential Problem}},\ }\href
  {https://doi.org/10.1103/PhysRev.94.1498} {\bibfield  {journal} {\bibinfo
  {journal} {Phys. Rev.}\ }\textbf {\bibinfo {volume} {94}},\ \bibinfo {pages}
  {1498} (\bibinfo {year} {1954})}\BibitemShut {NoStop}%
\bibitem [{\citenamefont {Mehl}\ and\ \citenamefont
  {Papaconstantopoulos}(1996)}]{Mehl1996}%
  \BibitemOpen
  \bibfield  {author} {\bibinfo {author} {\bibfnamefont {M.~J.}\ \bibnamefont
  {Mehl}}\ and\ \bibinfo {author} {\bibfnamefont {D.~A.}\ \bibnamefont
  {Papaconstantopoulos}},\ }\bibfield  {title} {\bibinfo {title} {Applications
  of a tight-binding total-energy method for transition and noble metals:
  Elastic constants, vacancies, and surfaces of monatomic metals},\ }\href
  {https://doi.org/10.1103/PhysRevB.54.4519} {\bibfield  {journal} {\bibinfo
  {journal} {Phys. Rev. B}\ }\textbf {\bibinfo {volume} {54}},\ \bibinfo
  {pages} {4519} (\bibinfo {year} {1996})}\BibitemShut {NoStop}%
\bibitem [{\citenamefont {Barreteau}\ \emph {et~al.}(2016)\citenamefont
  {Barreteau}, \citenamefont {Spanjaard},\ and\ \citenamefont
  {Desjonqu{\`{e}}res}}]{Barreteau2016}%
  \BibitemOpen
  \bibfield  {author} {\bibinfo {author} {\bibfnamefont {C.}~\bibnamefont
  {Barreteau}}, \bibinfo {author} {\bibfnamefont {D.}~\bibnamefont
  {Spanjaard}},\ and\ \bibinfo {author} {\bibfnamefont {M.-C.}\ \bibnamefont
  {Desjonqu{\`{e}}res}},\ }\bibfield  {title} {\bibinfo {title} {An efficient
  magnetic tight-binding method for transition metals and alloys},\ }\href
  {https://doi.org/https://doi.org/10.1016/j.crhy.2015.12.014} {\bibfield
  {journal} {\bibinfo  {journal} {C. R. Phys.}\ }\textbf {\bibinfo {volume}
  {17}},\ \bibinfo {pages} {406} (\bibinfo {year} {2016})}\BibitemShut
  {NoStop}%
\bibitem [{\citenamefont {Soulairol}\ \emph {et~al.}(2016)\citenamefont
  {Soulairol}, \citenamefont {Barreteau},\ and\ \citenamefont
  {Fu}}]{Soulairol2016}%
  \BibitemOpen
  \bibfield  {author} {\bibinfo {author} {\bibfnamefont {R.}~\bibnamefont
  {Soulairol}}, \bibinfo {author} {\bibfnamefont {C.}~\bibnamefont
  {Barreteau}},\ and\ \bibinfo {author} {\bibfnamefont {C.-C.}\ \bibnamefont
  {Fu}},\ }\bibfield  {title} {\bibinfo {title} {{Interplay between magnetism
  and energetics in Fe-Cr alloys from a predictive noncollinear magnetic
  tight-binding model}},\ }\href {https://doi.org/10.1103/PhysRevB.94.024427}
  {\bibfield  {journal} {\bibinfo  {journal} {Phys. Rev. B}\ }\textbf {\bibinfo
  {volume} {94}},\ \bibinfo {pages} {024427} (\bibinfo {year}
  {2016})}\BibitemShut {NoStop}%
\bibitem [{\citenamefont {Schena}(2010)}]{Schena2010}%
  \BibitemOpen
  \bibfield  {author} {\bibinfo {author} {\bibfnamefont {T.}~\bibnamefont
  {Schena}},\ }\emph {\bibinfo {title} {{Tight-Binding Treatment of Complex
  Magnetic Structures in Low-Dimensional Systems}}},\ \href
  {https://www.fz-juelich.de/SharedDocs/Downloads/PGI/PGI-1/EN/Schena_diploma_pdf.pdf}
  {\bibinfo {type} {Diploma thesis}},\ \bibinfo  {school} {TH Aachen} (\bibinfo
  {year} {2010})\BibitemShut {NoStop}%
\bibitem [{\citenamefont {Bruno}(2003)}]{Bruno2003}%
  \BibitemOpen
  \bibfield  {author} {\bibinfo {author} {\bibfnamefont {P.}~\bibnamefont
  {Bruno}},\ }\bibfield  {title} {\bibinfo {title} {Exchange interaction
  parameters and adiabatic spin-wave spectra of ferromagnets: A ``renormalized
  magnetic force theorem''},\ }\href
  {https://doi.org/10.1103/PhysRevLett.90.087205} {\bibfield  {journal}
  {\bibinfo  {journal} {Phys. Rev. Lett.}\ }\textbf {\bibinfo {volume} {90}},\
  \bibinfo {pages} {087205} (\bibinfo {year} {2003})}\BibitemShut {NoStop}%
\bibitem [{\citenamefont {Kvashnin}\ \emph {et~al.}(2016)\citenamefont
  {Kvashnin}, \citenamefont {Cardias}, \citenamefont {Szilva}, \citenamefont
  {Di~Marco}, \citenamefont {Katsnelson}, \citenamefont {Lichtenstein},
  \citenamefont {Nordstr\"om}, \citenamefont {Klautau},\ and\ \citenamefont
  {Eriksson}}]{Kvashnin2016}%
  \BibitemOpen
  \bibfield  {author} {\bibinfo {author} {\bibfnamefont {Y.~O.}\ \bibnamefont
  {Kvashnin}}, \bibinfo {author} {\bibfnamefont {R.}~\bibnamefont {Cardias}},
  \bibinfo {author} {\bibfnamefont {A.}~\bibnamefont {Szilva}}, \bibinfo
  {author} {\bibfnamefont {I.}~\bibnamefont {Di~Marco}}, \bibinfo {author}
  {\bibfnamefont {M.~I.}\ \bibnamefont {Katsnelson}}, \bibinfo {author}
  {\bibfnamefont {A.~I.}\ \bibnamefont {Lichtenstein}}, \bibinfo {author}
  {\bibfnamefont {L.}~\bibnamefont {Nordstr\"om}}, \bibinfo {author}
  {\bibfnamefont {A.~B.}\ \bibnamefont {Klautau}},\ and\ \bibinfo {author}
  {\bibfnamefont {O.}~\bibnamefont {Eriksson}},\ }\bibfield  {title} {\bibinfo
  {title} {{Microscopic Origin of Heisenberg and Non-Heisenberg Exchange
  Interactions in Ferromagnetic bcc Fe}},\ }\href
  {https://doi.org/10.1103/PhysRevLett.116.217202} {\bibfield  {journal}
  {\bibinfo  {journal} {Phys. Rev. Lett.}\ }\textbf {\bibinfo {volume} {116}},\
  \bibinfo {pages} {217202} (\bibinfo {year} {2016})}\BibitemShut {NoStop}%
\bibitem [{\citenamefont {Cardias}\ \emph {et~al.}(2017)\citenamefont
  {Cardias}, \citenamefont {Szilva}, \citenamefont {Bergman}, \citenamefont
  {Marco}, \citenamefont {Katsnelson}, \citenamefont {Lichtenstein},
  \citenamefont {Nordstr{\"o}m}, \citenamefont {Klautau}, \citenamefont
  {Eriksson},\ and\ \citenamefont {Kvashnin}}]{Cardias2017}%
  \BibitemOpen
  \bibfield  {author} {\bibinfo {author} {\bibfnamefont {R.}~\bibnamefont
  {Cardias}}, \bibinfo {author} {\bibfnamefont {A.}~\bibnamefont {Szilva}},
  \bibinfo {author} {\bibfnamefont {A.}~\bibnamefont {Bergman}}, \bibinfo
  {author} {\bibfnamefont {I.~D.}\ \bibnamefont {Marco}}, \bibinfo {author}
  {\bibfnamefont {M.~I.}\ \bibnamefont {Katsnelson}}, \bibinfo {author}
  {\bibfnamefont {A.~I.}\ \bibnamefont {Lichtenstein}}, \bibinfo {author}
  {\bibfnamefont {L.}~\bibnamefont {Nordstr{\"o}m}}, \bibinfo {author}
  {\bibfnamefont {A.~B.}\ \bibnamefont {Klautau}}, \bibinfo {author}
  {\bibfnamefont {O.}~\bibnamefont {Eriksson}},\ and\ \bibinfo {author}
  {\bibfnamefont {Y.~O.}\ \bibnamefont {Kvashnin}},\ }\bibfield  {title}
  {\bibinfo {title} {{The Bethe-Slater curve revisited; new insights from
  electronic structure theory}},\ }\href
  {https://doi.org/10.1038/s41598-017-04427-9} {\bibfield  {journal} {\bibinfo
  {journal} {Sci. Rep.}\ }\textbf {\bibinfo {volume} {7}},\ \bibinfo {pages}
  {4058} (\bibinfo {year} {2017})}\BibitemShut {NoStop}%
\bibitem [{\citenamefont {Thonig}\ and\ \citenamefont
  {Henk}(2014)}]{Thonig2014}%
  \BibitemOpen
  \bibfield  {author} {\bibinfo {author} {\bibfnamefont {D.}~\bibnamefont
  {Thonig}}\ and\ \bibinfo {author} {\bibfnamefont {J.}~\bibnamefont {Henk}},\
  }\bibfield  {title} {\bibinfo {title} {{Gilbert damping tensor within the
  breathing Fermi surface model: anisotropy and non-locality}},\ }\href
  {https://doi.org/10.1088/1367-2630/16/1/013032} {\bibfield  {journal}
  {\bibinfo  {journal} {New J. Phys}\ }\textbf {\bibinfo {volume} {16}},\
  \bibinfo {pages} {013032} (\bibinfo {year} {2014})}\BibitemShut {NoStop}%
\bibitem [{\citenamefont {Tung}\ and\ \citenamefont {Guo}(2011)}]{Tung2011}%
  \BibitemOpen
  \bibfield  {author} {\bibinfo {author} {\bibfnamefont {J.~C.}\ \bibnamefont
  {Tung}}\ and\ \bibinfo {author} {\bibfnamefont {G.~Y.}\ \bibnamefont {Guo}},\
  }\bibfield  {title} {\bibinfo {title} {Ab initio studies of spin-spiral waves
  and exchange interactions in $3d$ transition metal atomic chains},\ }\href
  {https://doi.org/10.1103/PhysRevB.83.144403} {\bibfield  {journal} {\bibinfo
  {journal} {Phys. Rev. B}\ }\textbf {\bibinfo {volume} {83}},\ \bibinfo
  {pages} {144403} (\bibinfo {year} {2011})}\BibitemShut {NoStop}%
\bibitem [{\citenamefont {T\"ows}\ and\ \citenamefont
  {Pastor}(2012)}]{Toews2012}%
  \BibitemOpen
  \bibfield  {author} {\bibinfo {author} {\bibfnamefont {W.}~\bibnamefont
  {T\"ows}}\ and\ \bibinfo {author} {\bibfnamefont {G.~M.}\ \bibnamefont
  {Pastor}},\ }\bibfield  {title} {\bibinfo {title} {Theoretical study of the
  temperature dependence of the magnon dispersion relation in transition-metal
  wires and monolayers},\ }\href {https://doi.org/10.1103/PhysRevB.86.054443}
  {\bibfield  {journal} {\bibinfo  {journal} {Phys. Rev. B}\ }\textbf {\bibinfo
  {volume} {86}},\ \bibinfo {pages} {054443} (\bibinfo {year}
  {2012})}\BibitemShut {NoStop}%
\bibitem [{\citenamefont {Solovyev}(2021)}]{Solovyev2020}%
  \BibitemOpen
  \bibfield  {author} {\bibinfo {author} {\bibfnamefont {I.~V.}\ \bibnamefont
  {Solovyev}},\ }\bibfield  {title} {\bibinfo {title} {Exchange interactions
  and magnetic force theorem},\ }\href
  {https://doi.org/10.1103/PhysRevB.103.104428} {\bibfield  {journal} {\bibinfo
   {journal} {Phys. Rev. B}\ }\textbf {\bibinfo {volume} {103}},\ \bibinfo
  {pages} {104428} (\bibinfo {year} {2021})}\BibitemShut {NoStop}%
\bibitem [{\citenamefont {Toepler}\ \emph {et~al.}(2021)\citenamefont
  {Toepler}, \citenamefont {Henk},\ and\ \citenamefont {Mertig}}]{Toepler2021}%
  \BibitemOpen
  \bibfield  {author} {\bibinfo {author} {\bibfnamefont {F.}~\bibnamefont
  {Toepler}}, \bibinfo {author} {\bibfnamefont {J.}~\bibnamefont {Henk}},\ and\
  \bibinfo {author} {\bibfnamefont {I.}~\bibnamefont {Mertig}},\ }\bibfield
  {title} {\bibinfo {title} {{Ultrafast spin dynamics in inhomogeneous systems:
  a density-matrix approach applied to Co/Cu interfaces}},\ }\href
  {https://doi.org/10.1088/1367-2630/abe72b} {\bibfield  {journal} {\bibinfo
  {journal} {New J. Phys}\ }\textbf {\bibinfo {volume} {23}},\ \bibinfo {pages}
  {033042} (\bibinfo {year} {2021})}\BibitemShut {NoStop}%
\bibitem [{\citenamefont {Hamamera}\ \emph {et~al.}(2022)\citenamefont
  {Hamamera}, \citenamefont {Guimar{\~a}es}, \citenamefont {dos Santos~Dias},\
  and\ \citenamefont {Lounis}}]{Hamamera2022}%
  \BibitemOpen
  \bibfield  {author} {\bibinfo {author} {\bibfnamefont {H.}~\bibnamefont
  {Hamamera}}, \bibinfo {author} {\bibfnamefont {F.~S.~M.}\ \bibnamefont
  {Guimar{\~a}es}}, \bibinfo {author} {\bibfnamefont {M.}~\bibnamefont {dos
  Santos~Dias}},\ and\ \bibinfo {author} {\bibfnamefont {S.}~\bibnamefont
  {Lounis}},\ }\bibfield  {title} {\bibinfo {title} {Polarisation-dependent
  single-pulse ultrafast optical switching of an elementary ferromagnet},\
  }\href {https://doi.org/10.1038/s42005-021-00798-8} {\bibfield  {journal}
  {\bibinfo  {journal} {Commun. Phys.}\ }\textbf {\bibinfo {volume}
  {5}},\ \bibinfo {pages} {16} (\bibinfo {year} {2022})}\BibitemShut {NoStop}%
\end{thebibliography}
